\documentclass[letterpaper,aps,preprintnumbers,longbibliography,notitlepage]{revtex4-1}
\usepackage[latin9]{inputenc}
\usepackage{wrapfig}
\usepackage{amsmath}
\usepackage{amssymb}
\usepackage{graphicx}
\usepackage{esint}

\makeatletter

\pdfpageheight\paperheight
\pdfpagewidth\paperwidth

\makeatother

\begin{document}

\preprint{FERMILAB-PUB-15-268-A}

\title{Statistical Measures of Planck Scale Signal Correlations in Interferometers}

\author{Craig J. Hogan}

\affiliation{University of Chicago and Fermilab Center for Particle Astrophysics}

\author{Ohkyung Kwon\footnote{E-mail: o.kwon@kaist.ac.kr}}

\affiliation{University of Chicago and Korea Advanced Institute of Science and
Technology}
\begin{abstract}
A model-independent statistical framework is presented to interpret
data from systems where the mean time derivative of positional cross
correlation between world lines, a measure of spreading in a quantum
geometrical wave function, is measured with a precision smaller than
the Planck time. The framework provides a general way to constrain
possible departures from perfect independence of classical world lines,
associated with Planck scale bounds on positional information. A parameterized
candidate set of possible correlation functions is shown to be consistent
with the known causal structure of the classical geometry measured
by an apparatus, and the holographic scaling of information suggested
by gravity. Frequency-domain power spectra are derived that can be
compared with interferometer data. Simple projections of sensitivity
for realistic experimental set-ups suggests that measurements will
 confirm or rule out a class of Planck scale
departures from classical geometry. 
\end{abstract}
\maketitle

\section{Introduction}

Although it is widely believed that the world follows quantum principles,
it is not known how classical dynamical space-time emerges from a
quantum mechanical system \cite{Rovelli2004,Thiemann2008,Ashtekar2012}.
Standard field theory, often regarded as the basis of a fundamental
quantum theory, encounters well-known, unavoidable divergences at
the Planck scale \cite{DeservanNieuwenhuizen1974,Weinberg1996,Wilczek1999}.
The divergences may be avoided in a theory based on other fundamental
objects such as strings\cite{EllisMavromatosNanopoulos1992,Polchinski1998,Hossenfelder2013},
loops or spin networks\cite{Rovelli2004,Thiemann2008,Ashtekar2012},
and/or noncommutative geometries\cite{DouglasNekrasov2001}; these
theories can even lead to predictions of new behavior at the Planck
scale, such as quantization of volume or area. However, for large
systems, many issues remain unresolved, for example, the foundational
inconsistency of quantum theory with the ``local realism'' associated
with classical geometrical paths and events\cite{GiddingsMarolfHartle2006,Giddings2007,Banks2009,Banks2011},
the holographic encoding of information in dynamical gravitational
systems, particularly black holes\cite{BardeenCarterHawking1973,Bekenstein1973,Bekenstein1974,Hawking1974,Hawking1975,tHooft1993,Jacobson1995,Susskind1995,Bousso2002,Verlinde2011},
and unphysically large masses of some field systems in large volumes\cite{CohenKaplanNelson1999,Hogan2014}.

Most theoretical attention has concentrated on the ultraviolet part
of the problem--- the Planck scale regime dominated by string theory.
On the other hand, the large-scale or infrared paradoxes also presumably
arise because field theory is quantized on a fixed classical background,
whereas the real space-time system emerges, along with the classical
notion of locality, from a quantum system based on a Planckian bandwidth
limit or discrete structure. While ultraviolet effects at the Planck
scale cannot be measured directly, scaling from standard quantum mechanics
suggests that indirect effects of Planck scale physics may be detectable
in certain kinds of measurements in laboratory systems via spreading
of geometrical position states over macroscopic distance. These arguments 
are reviewed below in section \ref{sub:General-Framework-Scaling}.

Correlations signifying departures from perfect classical space could
plausibly appear, according to standard quantum mechanics, on a scale
given by diffraction of Planck frequency radiation propagating across
an apparatus, or equivalently, the quantum position uncertainty of
a Planck mass particle over a duration corresponding to an apparatus
light-crossing time \cite{Hogan2008a,Hogan2008,Hogan2012,KwonHogan2014,Hogan2014}.
Field theory assumes a classically coherent background geometry, and
predicts that such nonlocal quantum correlations of the background
should be negligible.

Although the approximate magnitude of the correlations can be guessed
from scaling, their detailed character is not known. In the absence
of a standard theory for large systems in emergent geometry, in this
paper we take an empirical, experimental approach to the problem.
We ask, what can precise laboratory measurements of the relative positions
of massive bodies at rest tell us about the character of positional
information contained in the space-time itself, and in particular,
the detailed degrees of freedom of the quantum system that gives rise
to space and locality? We focus on what can be learned about the emergence
of classical positions in space from the best available technique
for nonlocal measurement of position, interferometry\cite{Adhikari2014}.

The predictions of any quantum theory are expressed in terms of correlations
between observables. In the case of a space-time built from a quantum
system, departures from classicality should appear in time-averaged
correlations of positions of massive bodies. For a given configuration
of massive bodies--- for example, the mirrors in an interferometer---
the forms that the correlation functions can take are constrained
by general principles of symmetry and causality of the emerged space.
We argue here that possible forms can be classified and evaluated
from the known structure of a macroscopic apparatus, without knowing
the elements and dynamics of the underlying quantum space-time degrees
of freedom.

As one example of an observable correlation, signals from a pair of
interferometers can be combined into a correlation function, whose
mean time derivative has the dimension of time. If that has a value
of the order of the Planck time, the information content of the measured
spatial relationships is comparable with the total information content
suggested by covariant holographic bounds on entropy \cite{Hogan2008a,Hogan2008,Hogan2012,Hogan2012b,KwonHogan2014}.
Interferometers are now capable of achieving measurements of coherence
with Planck precision\cite{AffeldtEtAl2014,DooleyLIGO2015,ChouEtal2015,ChouEtal:2016,LIGO2015a,LIGO2016},
so the framework presented here can be used to interpret their signals
in terms of Planck scale bounds on information.

The goal is to use data to test a general hypothesis: that geometrical
information about directions between world lines in space, defined
by measurements of the position of massive bodies with electromagnetic
waves, obeys Planck scale covariant constraints on information. With
the framework presented here, statistical properties of data in a
broad class of experiments can be compared with each other, and used
to test specific forms of this general hypothesis, even in the absence
of a specific fundamental theory. Experiments can explore and constrain
any form of Planckian correlations in space-time position consistent
with symmetries imposed by the causal structure of the emergent system.

We apply our framework here to a particular kind of apparatus, a correlated
pair of Michelson interferometers, such as the Fermilab Holometer\cite{ChouEtal2015,ChouEtal:2016},
and also discuss how it can be generalized. In previous work\cite{Hogan2008a,Hogan2008,Hogan2012,Hogan2012b,KwonHogan2014},
specific models of holographic spatial position states were used to
predict properties of measurable ``holographic noise'' in interferometers.
The more general framework here allows a broader class of quantum
models to be tested, even some with no corresponding classical interpretation.
In all cases, the coherence is characterized by a key universal parameter:
the time derivative of a measured correlation function, normalized
to the size of an apparatus. We propose a scheme that spans the full
range of possible models, characterized by two universal parameters
of the order of the Planck time that can be directly constrained by
data.

Of course, different experimental set-ups, with different causal relationships
between the world lines of their optical elements, will constrain
Planckian departures from classical coherence in different ways. We
do not exhaust all those possibilities here, but general principles
of this framework can be extended to compare them with each other
and with predictions of proposed theories. For example, the Michelson
systems mainly considered here (and used in the initial stage of the
Holometer) measure only correlations spatially related by shear or
strain transformations; they do not measure the important independent
modes corresponding to pure rotations\cite{Hogan2015,HoganKwonRichardson2016}. Those modes,
and apparatus that could measure them in a future experimental program
at the Holometer, will be treated in upcoming analyses.

\section{General Framework\label{sec:General-Framework}}

\subsection{Scaling of Geometrical Correlations\label{sub:General-Framework-Scaling}}

\begin{wrapfigure}{o}{0.215\columnwidth}%
\begin{centering}
\includegraphics[scale=0.325]{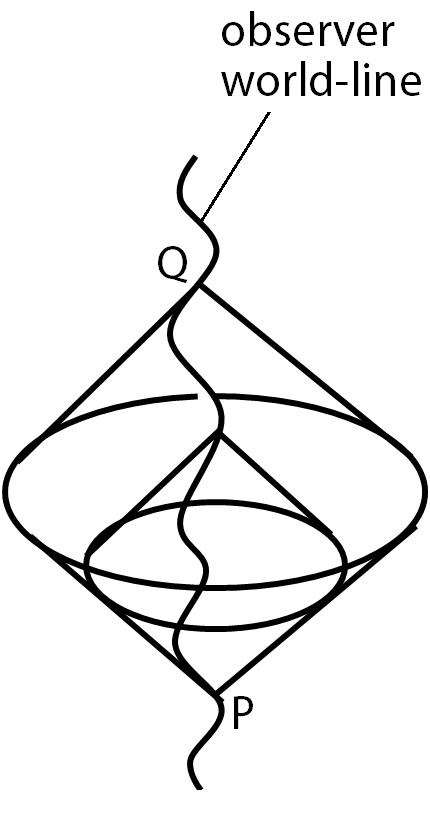} 
\par\end{centering}

\protect\caption{Causal diamonds.\label{fig:Causal-diamonds}}
\end{wrapfigure}%

Causal diamonds are defined as the 4-volume of spacetime over which
an observer (defined by a timelike world-line) can do experiments.
It is bounded by two null light-cones connecting two points on the
observer's world-line that are timelike separated (see Figure \ref{fig:Causal-diamonds}).
A world line defines a causal diamond associated with any proper time
interval of duration $\tau$. It has been proposed that nested sequences
of causal diamonds--- covering increasing intervals along a timelike
trajectory--- correspond to sequences of nested Hilbert spaces \cite{Banks2009,Banks2011}.
These causal diamonds serve as reference boundaries for geometrical
information.

Consider an observable operator $\hat{x}$, with the dimension of
length, that characterizes a set of world lines in some extended 4D
volume of flat space-time. In general, a measurement of $\hat{x}$
requires nonlocal propagation of light within a causal diamond; it
represents a projection of an extended 4D quantum system. In the interferometer
systems considered here, an observable signal depends on the world
lines of an arrangement of mirrors, and encodes nonlocal information
about their positions and about the state of the laser field. Here,
we ignore these standard quantum degrees of freedom; the operators
$\hat{x}$ measure new noncommuting degrees of freedom that differ
from the standard motion\cite{SaleckerWigner1958} of a massive body
in a classical geometry. A measurement of $\hat{x}$ represents a
measurement of positional degrees of freedom that are assumed to be
classical in the standard quantization of fields and optical elements
in an interferometer\cite{Caves1980,CavesThorneDreverSandbergZimmermann1980}.
They correspond to new degrees of freedom associated with the nonclassical
character of emergent world lines; thus, we know neither the Hamiltonian
nor the conjugate operators for $\hat{x}$. (Other forms of entanglement
with Planck scale geometrical states have been discussed in ref. \cite{RuoBercheraDegiovanniOlivaresGenovese2013}.)
The value of $\hat{x}$ does not depend on the masses or other properties
of bodies, only on their relative positions.

Suppose that the measured value of $\hat{x}$ departs by some amount
$\Delta x=\hat{x}-\bar{x}$ from its expected value $\bar{x}$ in
classical space-time. A measure of the deviation from perfect classical
coherence is given by a time-domain correlation function of the form,
\begin{equation}
\Xi(\tau)\equiv\langle\Delta x(t)\Delta x(t+\tau)\rangle_{t}F(\tau).\label{DeltaCorrelation}
\end{equation}
Here, $F(\tau)$ denotes a projection determined by the configuration
of a measurement apparatus. In general, correlations can be measured
for two world lines, say $\hat{x}_{A}$ and $\hat{x}_{B}$; most of
this paper focuses on the configurations where $A$ and $B$ causal
diamonds are almost the same.

The positional correlation encoded in $\Xi$ measures a lack of independence
of values of $\hat{x}$. In a classical space-time, $\Xi$ vanishes;
this is possible because the information density in a continuum is
infinite. The same is true in field theory, which assumes a classical
continuous space-time. In an emergent quantum space-time, $\Xi$ in
general does not vanish, and the positions of world-lines in some
measurements can decohere gradually with time. The 2D density of position
eigenstates--- that is, the number of independent world-line eigenstates
per 2-volume in a particular $\hat{x}$ projection--- has a finite
value given by $\Xi^{-1}$. Its value depends on the projection defined
by the measurement, and the invariant classical positional relationships
between the world lines, particularly the causal diamonds traced by
light propagation in the apparatus.

The scaling with $\tau$ can be estimated from dimensional considerations,
or from analogy with other, standard quantum systems. It is plausible
that world lines decohere slightly at long durations, by an amount
that would not yet have been detected. As one example of a quantum
system with this behavior, consider the standard position wave function
for the state of a particle of mass $m$ at rest, that lasts for duration
$\tau$. This form of standard Heisenberg uncertainty can be derived
in several ways\cite{Hogan2008a,Hogan2008,Hogan2012,Hogan2014}, for
example from a nonrelativistic Schr\"odinger equation for a mass
$m$, in a path integral approach by extremizing the action of a particle
or body whose motion is described by a wave equation with de Broglie
wavelength $\hbar/mc$, or from a Wheeler-De Witt equation for a pendulum
of mass $m$ in the low frequency, nearly free particle limit. For
all of these systems, with the standard assumption that directions
in space are independent, the wave function in classical position
and time obeys 
\begin{equation}
(\partial_{i}\partial^{i}-2i(m/\hbar)\partial_{t})\psi=0.\label{newparax}
\end{equation}
This equation can be solved with pure (infinite) plane wave eigenmodes.
However, the state of a body localized in space, whose position is
prepared and measured at two times separated by an interval much longer
than the inverse de Broglie frequency, is better described by a symmetric
gaussian solution for $\psi(r,t)$, where $r^{2}=x_{i}x^{i}$, with
a probability density on constant-time surfaces given by 
\begin{equation}
|\langle\psi^{*}|\psi\rangle|^{2}\propto e^{-r^{2}/\sigma(t)^{2}},\label{probdense}
\end{equation}
whose variance $\sigma^2$ depends on the preparation of the state.

\begin{figure}
\includegraphics[scale=0.18]{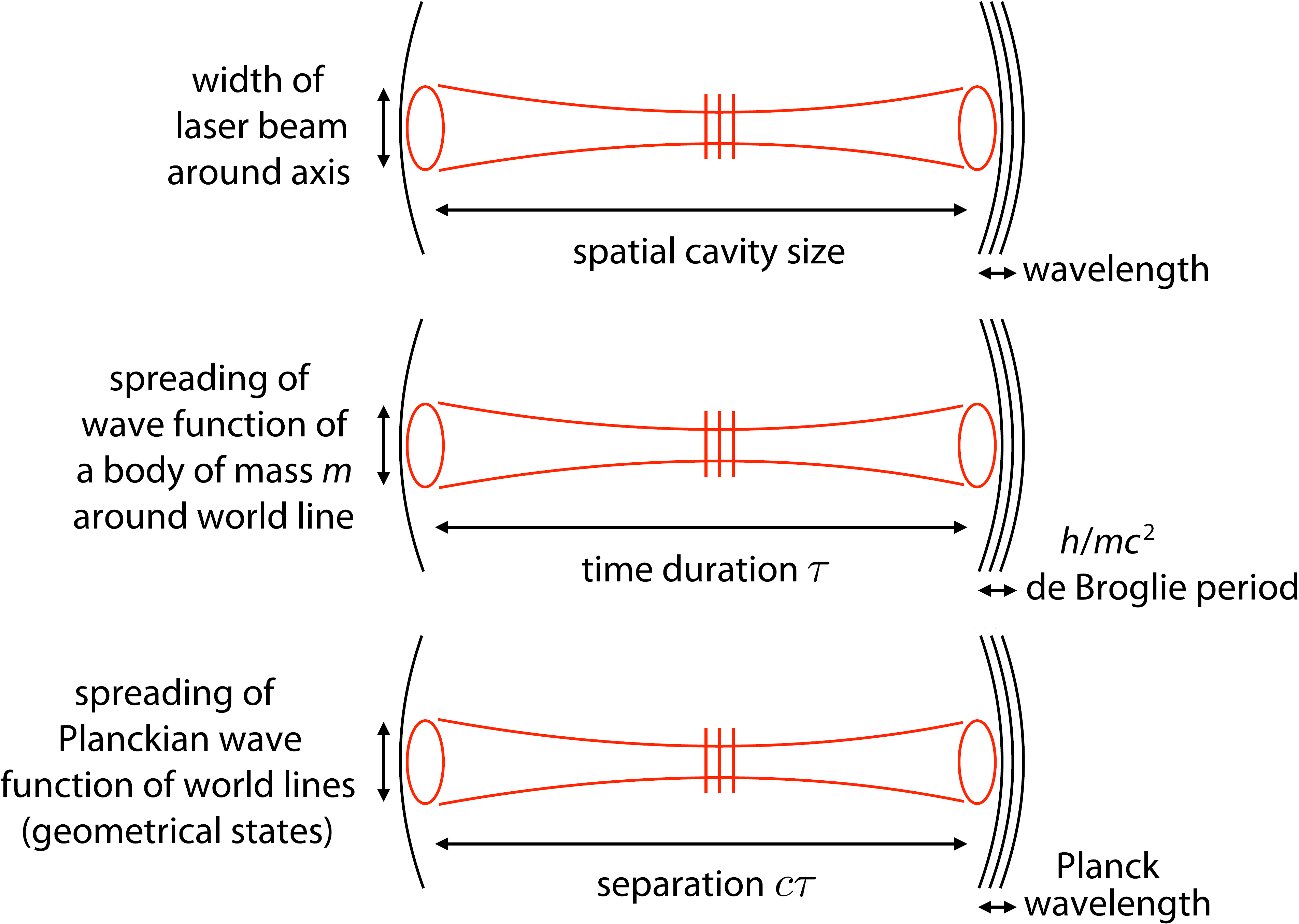}

\protect\caption{Paraxial solutions and ``world tubes.'' \label{fig:Paraxial}}

\end{figure}

These ``paraxial'' solutions for matter de Broglie waves are mathematically
the same as the standard normal modes of light\cite{Siegman1986lasers}
in a laser cavity (see Figure \ref{fig:Paraxial}). In the matter
system, $t$ takes the place of the laser beam axis coordinate $z$
(that is, constant-phase de Broglie wavefronts are nearly surfaces
of constant time), and position measurements at particular times,
over an interval much smaller than the interval between measurements,
takes the place of thin mirror surface boundary conditions \cite{Hogan2014}.
In this family of solutions, the rate of spatial spreading depends
on the preparation of a state: $d\sigma/dt$ is smaller for larger
$\sigma$. In the same way that a larger beam waist leads to a smaller
dispersion angle, a larger position uncertainty leads to slower spreading.
Spreading of the wave function with time is unavoidable; any solution
of duration $\tau$ has a mean variance of position at least as large
as 
\begin{equation}
\langle(r(t+\tau)-r(t))^{2}\rangle=\sigma_{0}^{2}(\tau)>\hbar\tau/2m.\label{minimum}
\end{equation}
Thus, $d\sigma_{0}/d\tau=\hbar/2m$ gives the minimum rate of spreading
in the position state of a body at rest over time.

Of course, these states also have an associated indeterminacy of momentum.
Although the particle is classically ``at rest'', and there is a
well defined expectation value for its rest frame, the actual velocity
is indeterminate. In this situation, the position variance between
two times cannot be ``squeezed away'' into momentum uncertainty,
or vice versa; Eq. (\ref{minimum}) gives the minimum spread of position
between two times for \textit{any} state \cite{Caves1980,CavesThorneDreverSandbergZimmermann1980,Gardiner2004quantum}.
The momentum uncertainty in the minimum-position-uncertainty state
decreases with $\tau$ like $\Delta p\approx(2m\hbar/\tau)^{1/2}$,
which is why de Broglie wave states behave almost like classical world
lines on large scales.

Similarly, if paths in a background quantum space time are described
by quantum wave propagation similar to Eq. (\ref{newparax}), with
a fundamental timescale set by the Planck time (that is, setting $m=m_{P}=\hbar/c^{2}t_{P}$),
then its emergent world line states spread over duration $\tau$ by
about 
\begin{equation}
\Xi(\tau)\approx\langle(x(t+\tau)-x(t))^{2}\rangle\approx c^{2}\tau t_{P}/2.
\end{equation}
To be sure, this background spreading cannot be detected by a local
measurement, in the same way that quantum decoherence between independent
paths of mass $m_{P}$ bodies can be measured by comparing their positions,
because the geometrical states of nearby world lines are entangled
with each other. It may also include exotic correlations between different
directions and components of position, which we have not included.
However, the scaling of $\sigma_{0}^{2}(\tau)$ and $\Xi(\tau)$ with
$\tau$ are the same: world lines should decohere at a rate given
approximately by $d\Xi/d\tau\approx c^{2}t_{P}/2$.

Figure \ref{fig:Paraxial} shows the mathematical parallels. The spreading
of the wave function of a body of mass $m$ around its world line
over a duration $\tau$ (which we can loosely think of as ``world
tubes'') is seen as mathematically similar to the spreading of a
Planckian wave function describing emergent world line states over
a separation $c\tau$, once the fundamental scale is set via the Planck
mass. These ``world tubes'' can now describe the decoherence of
geometrical states.

For large systems (that is, $\tau\gg t_{P}$), the variance of position
is much larger than the typical value $\Xi\approx c^{2}t_{P}^{2}$
from Planck scale field fluctuations on a classical background. At
the same time, the variance is very small, and the trajectories behave
nearly classically. Because the deviation can only be seen in a highly
precise nonlocal measurement of correlations, it would not have been
detected in previous experiments. A 100m laboratory scale experiment
is $\approx10^{37}$ Planck lengths across; Planckian spreading on
the corresponding timescale ($10^{37}t_{P}\approx10^{-6}$ sec) is
only $\approx10^{18.5}ct_{P}\approx10^{-16.5}$ meters, and the typical
variation in rest frame velocity is about $\approx10^{-18.5}c$, or
a few mm per year--- an order of magnitude slower than continental
drift.

The same scaling is suggested by applying holographic bounds to information
in the relative position of world lines. The holographic principle
posits covariantly that the information ${\cal I}$ in a causal diamond
of duration $\tau$ is one quarter of the area of its bounding surface
in Planck units, or ${\cal I}_{H}=(\pi/4)(\tau/t_{P})^{2}$ in flat
space. Suppose that the position operator $\hat{x}$ depends on positions
distributed over a 3-volume $\approx(c\tau)^{3}$, and that the density
of information in the radial dimension of the causal diamond has some
value $\ell_{r}^{-1}$ in its rest frame, independent of system size.
The number of position states is then ${\cal I}_{x}\approx(c\tau)^{3}\ell_{r}^{-1}\Xi^{-1}$.
If we require that ${\cal I}_{x}<{\cal I}_{H}$, so that the holographic
information bound applies to spatial information, we obtain 
\begin{equation}
\Xi(\tau)>\tau(c^{3}t_{P}^{2}/\ell_{r})(\pi/4).
\end{equation}
Thus, there is some universal value of $\dot{\Xi}\equiv d\Xi/d\tau$
associated with holographic scaling of position information. Since
the natural value of $\ell_{r}$ is of order $ct_{P}$, the natural
value for $\dot{\Xi}$ is again of order $c^{2}t_{P}$.

The value of $\dot{\Xi}$ depends on the configuration of the world
lines represented in the operator $\hat{x}$. For some $\hat{x}$
operators that do not measure degrees of freedom where the information
is bounded, it may vanish by symmetry; that is, the uncertainty could
be squeezed into unmeasured degrees of freedom. Although the analogy
(Eq. \ref{newparax}) is suggestive of how solutions should scale,
we do not know how measurements in different directions, or between
different world lines, should relate to each other, or how the wave
function should be interpreted physically. The experimental program
is to measure correlations where we can, and perhaps uncover clues
to the underlying theory. 

The current exercise is to lay out statistical measures that can measure
the value of $\dot{\Xi}$ with data from simple experimental setups.
The form of $\Xi$ can be constrained from general considerations
about the causal diamonds corresponding to the arrangement of mirrors
in an apparatus, so well-characterized universal conclusions about
quantum geometry can be drawn from a specific experimental result.

\subsection{Constraints on Correlations in Interferometer Signals\label{sub:Constraints}}

Consider two Michelson interferometers, $A$ and $B$, with dark port
signals $x_{A}(t)$ and $x_{B}(t)$ calibrated in length units. The
statistical quantity estimated from a time stream of data is a cross-correlation
between interferometers $A$ and $B$: 
\begin{equation}
\Xi(\tau)\equiv\langle x_{A}(t)x_{B}(t+\tau)\rangle.
\end{equation}
The projection $F(\tau)$ (that is, the connection between the signal
$x$ and $\Delta x$ in Eq. \ref{DeltaCorrelation}) has here been
absorbed into the definition of the apparatus that produces the signal.

Classically, the signal $x$ sees the departure of the length difference
between the two arms from the average value. In a static classical
space-time with no physical coupling between the two interferometers
and no gravitational waves, the average cross correlation vanishes,
$\Xi=0$. However, if the two interferometers share a volume of space-time---
that is, if the causal diamonds traced by the propagation of light
in their arms overlap in some way--- a correlation might be induced
by Planckian holographic bounds on the amount of positional information
in the system measured by the two sets of optics. They may be entangled
by geometrical states in a way that statistically correlates the measurements.

We now propose a simple framework to test the general holographic
noise hypothesis: that there is a nonzero correlation between $A$
and $B$ due to entanglement of the emergent space measured by the
two interferometers. This framework is formulated in a general way
so that the statistical results of a specific experiment can be used
to constrain any (future) candidate universal theory of quantum geometry.

\begin{figure}
\includegraphics[scale=0.3]{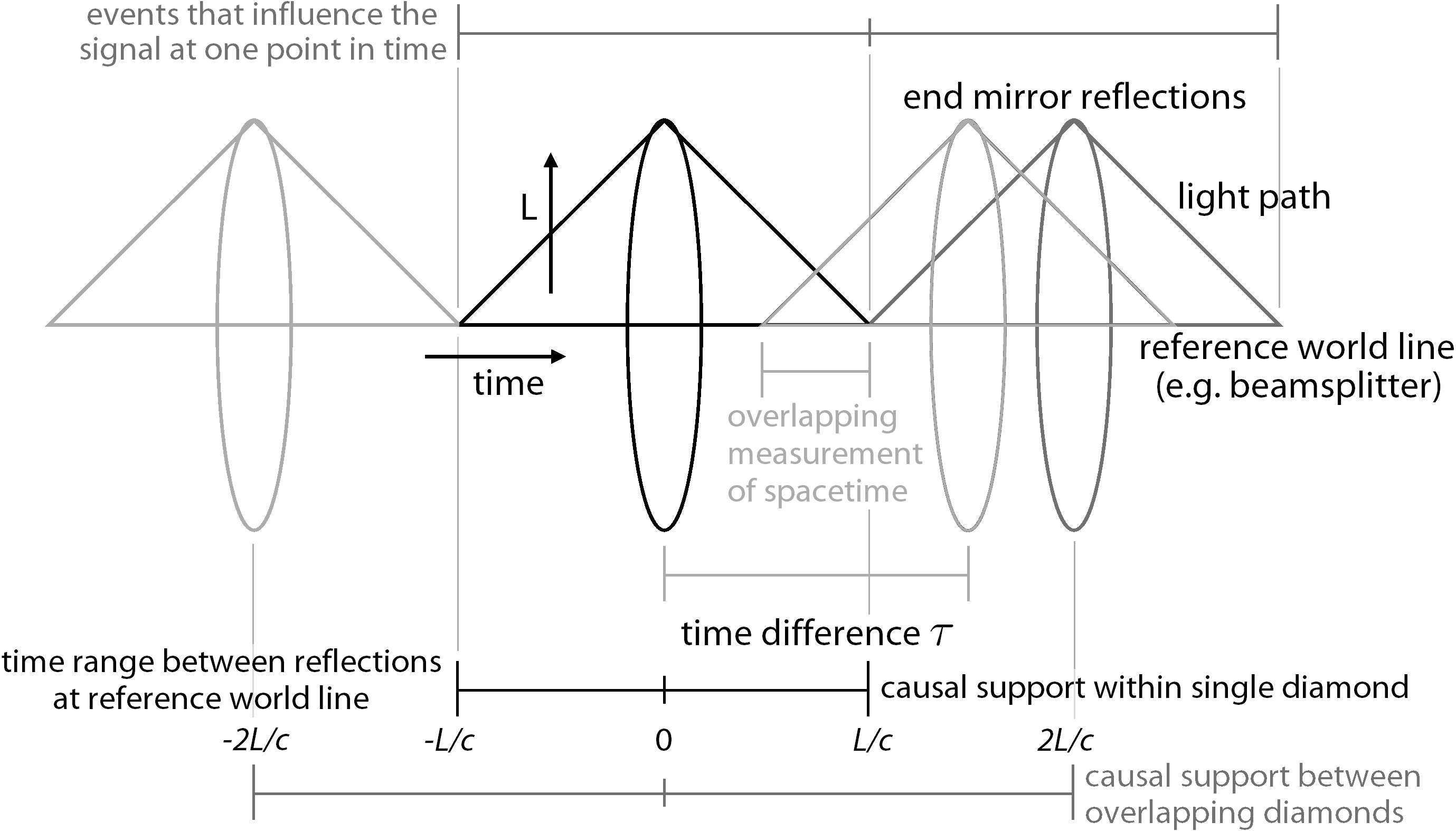}

\protect\caption{A comparison of causal support in the two classes of correlation functions.\label{fig:causalcomp}}
\end{figure}

Even without a specific fundamental theory, we know that holographic
correlations (if they exist at all) are subject to a few simple mathematical
and physical constraints. 
\begin{enumerate}
\item $\Xi(\tau)$ must be symmetric, $\Xi(\tau)=\Xi(-\tau)$. Time symmetry
and general covariance dictate that $\Xi(\tau)$ must be either symmetric
or antisymmetric. However, Parseval's theorem requires that $\Xi(\tau)$
must have a nonzero value at zero lag if there is to be any measurable
fluctuation in a real-valued signal, eliminating the latter choice. 
\item The correlation function must respect the causal region in the emergent
space. For a simple configuration of two co-aligned Michelson interferometers
with characteristic scale $L$ (e.g. arm length), we expect the functional
support to be limited by the causal diamond to the interval $\tau=(-L,+L)$
between reflections at the reference world line (e.g. the beamsplitter),
or the interval $\tau=(-2L,+2L)$ of events that influence the signal
at any given time (see Figure \ref{fig:causalcomp}). More general
interferometer configurations may add additional scales, or even additional
world-lines, and yield different functions. 
\item The hypothesis being tested is that relative positions of massive
bodies as encoded in $\Xi(\tau)$  are subject to Planck scale spreading
of information just described. For shear modes (but not rotation), scaling symmetry then requires that
where it does not vanish, $\Xi(\tau)$ is linear in $\tau$, with
a derivative $d\Xi/d\tau$ that does not depend on the size of the
apparatus (see nested causal diamond in Figure \ref{fig:timedomain-l/c}), but only its shape.
To match holographic information content, its absolute value is approximately
the Planck time $t_{P}$. A more careful evaluation of numerical factors
follows below. 
\end{enumerate}
With these constraints, for a given apparatus, candidate forms for
$\Xi(\tau)$ are completely determined by two parameters. We consider
below possible forms for $\Xi(\tau)$ which span the range of consistent
options. We also consider how the form of $\Xi(\tau)$ depends on
the apparatus configuration--- the sizes and relative positioning
of two interferometers in space. The detailed worked examples here
apply to configurations where the interferometers are adjacent and
almost co-located, so that causal diamonds substantially overlap;
the simple arrangement reduces the number of parameters needed to
characterize the system.

\section{Parameterized Models of Correlation Functions}

A broad class of correlation functions respects the constraints just
outlined. The models are described by two parameters that are universal
physical constants, that according to the holographic noise hypothesis
should not depend on apparatus size. Two basis vectors, along with
upper and lower bounds on normalization, appear to span the space
of possibilities consistent with the above constraints.

Following the two viable choices for causal support demonstrated in
Figure \ref{fig:causalcomp}, we first consider a class in which the
correlation function goes to zero at the lag corresponding to the
edge of the causal diamond defined by the end mirrors. We then present
a more general class in which the causal support contains all events
that influence the signal at a given time, of which the previous class
is a subset.

\subsection{Simple Case: Models Spanning $\pm L$ }

In this simplest class of models the correlation is required to drop
to zero at $\pm L$, the limits of functional support defined by a
single causal diamond. The linearity requirement from scaling symmetry
allows us to uniquely specify the correlation function by at most
two parameters: the zero lag value normalized to system scale (e.g.
arm length),
\begin{equation}
\xi_{0}\equiv\Xi(\tau=0)/L\label{eq:xi_def}
\end{equation}
and the norm of the derivative $|\dot{\Xi}|$. In other words,
\begin{equation}
\Xi(\tau)=\begin{cases}
\xi_{0}L-|\dot{\Xi}|\cdot\left|\tau\right| & \left|\tau\right|<\frac{L}{c}\\
0 & \left|\tau\right|>\frac{L}{c}
\end{cases}
\end{equation}

According to the holographic noise hypothesis, both $\xi_{0}$ and
$|\dot{\Xi}|$ are of the order of the Planck time. A choice of $\xi_{0}$
and $|\dot{\Xi}|$ predicts a frequency spectrum that can be compared
with Holometer data. A set of data can be used to generate a likelihood
function for the universal numbers.

\begin{figure}
\includegraphics[scale=0.8]{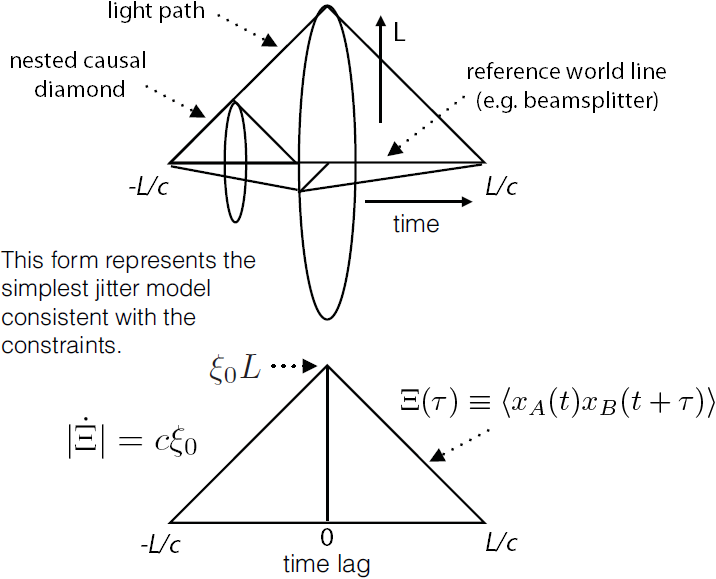}

\protect\caption{The simplest model, spanning $\pm L$: Linear time domain cross spectra.\label{fig:timedomain-l/c}}
\end{figure}

The parameter space formed by this class of models is actually one-dimensional,
because the derivative is constrained to be (see Figure \ref{fig:timedomain-l/c}):
\begin{equation}
|\dot{\Xi}|=c\xi_{0}
\end{equation}
reducing the two universal numbers into one parameter that sets the
normalization. This simple space is a subset of a more general space
of models that allows causal support ranging up to $\pm2L$, as we
will see in the next section.

This minimal form of holographic noise can be physically interpreted
as a new kind of exotic ``jitter'' in space and in the position
of reference world lines such as the beamsplitter. It was considered
as an illustrative model in previous work\cite{KwonHogan2014}, and
used to estimate experimental sensitivity for design purposes.

The fixed value of the derivative might seem overly constraining---
the causal diamond support constraint only sets a lower bound on $|\dot{\Xi}|$
at $|\dot{\Xi}|\geq c\xi_{0}$. We should not discount \textit{a priori}
the possibility of steeper slopes, in which case $\Xi(\tau)$ goes
negative (with discontinuities at $\pm L$), reflecting more exotic
nonlocal correlations in the causal 4-volume. This set of models merely
generalizes the simple model of jitter above, and could perhaps be
considered physically viable for holographic noise, as long as the
integral $\int d\tau\Xi$ remains nonnegative--- a condition satisfied
by $|\dot{\Xi}|\leq2c\xi_{0}$, an upper bound at which there is no
DC power in the frequency spectrum because the integral vanishes.

However, a statistical argument eliminates this freedom. For interferometers
that are adjacent and almost co-located, the cross-correlation $\Xi(\tau)$
can be thought of as an autocorrelation, since the causal diamonds
are almost the same and the geometrical states being measured are
substantially entangled. So $\Xi(\tau)$ is an autocorrelation of
a real-valued observable, the antisymmetric port beam power (calibrated
in length units), undergoing a wide-sense stationary random process.
This is possible if and only if the transform $\tilde{\Xi}(f)$ is
real, even, and nonnegative at all frequencies \cite{OppenheimVerghese2015}.
This consideration precludes the viability of models with slopes different
from $|\dot{\Xi}|=c\xi_{0}$.

The analytic form of the frequency domain power spectrum $\tilde{\Xi}(f)$
demonstrates this point:
\begin{eqnarray}
\tilde{\Xi}(f) & = & 2\int_{0}^{L/c}(\xi_{0}L-|\dot{\Xi}|\tau)\cos(2\pi f\tau)\textrm{d}\tau \nonumber \\
 & = & \frac{2|\dot{\Xi}|}{(2\pi f)^{2}}\left[1-\cos\left(\frac{f}{c/2\pi L}\right)\right]-\frac{2}{2\pi f}\left(\frac{L}{c}\right)\left(|\dot{\Xi}|-c\xi_{0}\right)\sin\left(\frac{f}{c/2\pi L}\right) \nonumber \\
 & = & \left(\frac{L}{c}\right)^{2}\left[|\dot{\Xi}|\textrm{sinc}^{2}\left(\frac{f}{c/\pi L}\right)-2\left(|\dot{\Xi}|-c\xi_{0}\right)\textrm{sinc}\left(\frac{f}{c/2\pi L}\right)\right]
\end{eqnarray}

\subsubsection*{Normalization from Holographic Gravity}

A fundamental theory of emergent geometry should make an exact prediction
for the parameters in correlation functions of a given apparatus. In the absence of such a theory,  it is useful to have concrete theoretical boundaries where one
can reasonably consider a particular hypothesis confirmed or ruled
out. 
We adopt a specific benchmark value as sketched in the appendix. The total number of states within a distance $L$ from
holographic or ``entropic'' gravity is equated with the number of states in a  simple noncommutative geometry, a quantum spin system of the same radius. The spin  algebra then provides an exact  numerical value
for the  variance of a spatial wave function in physical
length units at separation $L$: 
\begin{equation}
\langle x_{\perp}^{2}\rangle_{L}/L=\ell_{P}\equiv ct_{P}/\sqrt{4\pi}=\sqrt{\hbar G/4\pi c^{3}}=4.558\times10^{-36}{\rm m}.\label{eq:norm}
\end{equation}

This value sets a benchmark scale of the signal correlation function,
$\Xi(\tau=0)$, that saturates the holographic bound in a causal diamond
of radius $L$. The benchmark scale is to be used for guidance purposes
only. The validity of the normalization is independent from the general
framework presented throughout the rest of this paper, which relies
only on dimensional arguments and first principles of causal consistency
and symmetry.

The normalization scheme is valid only for the specific mode of correlation
being tested--- for the initial stage of the Holometer experiment,
these are \textit{shear} modes. The hypothesis being tested is that
geometrical information about \textit{directions} between world lines
in space obeys Planck scale covariant constraints, which potentially
includes shear and rotational modes, but the latter does not follow
the same causal structure used for this particular scheme, as will
be discussed in upcoming work. The simple Michelson systems considered
here are sensitive to strain fluctuations as well as shear ones, but
the number of quantum geometric degrees of freedom might scale differently
for strain correlations, and the predicted fluctuation amplitudes
do not necessarily reach detectable levels when constrained by holographic
bounds on information \cite{NgDam2000,Ng2002,KwonHogan2014}. We consider
such strain modes a less viable hypothesis, however, as a coherent
and directionally isotropic strain uncertainty likely violates general
covariance.

\subsubsection*{Normalization for Correlated Interferometer Noise}

Equation (\ref{eq:norm}) gives a reference scale for the transverse
variance associated with distance $L$, but in order to predict the
signal measured in interferometric experiments under this hypothesis,
we must conduct a holistic analysis of a quantum-geometrical system
of matter and light. It requires a full quantum theory that integrates
the Hilbert space of photon states with those of emergent-geometric
position states for the mass elements that comprise the optics. In
the absence of such a theory, we will establish upper and lower ranges
for $\xi_{0}$ (and therefore also for $|\dot{\Xi}|$).

A benchmark for $\xi_{0}$ is when Eq. (\ref{eq:xi_def}) and (\ref{eq:norm})
are compared straightforwardly, i.e. $\xi_{0}=\ell_{P}\equiv ct_{P}/\sqrt{4\pi}$.
These describe models that match holographic position uncertainty
(and holographic slope, $|\dot{\Xi}|=c\ell_{P}$).

An upper estimate for $\xi_{0}$ is posited when the measurements
of two orthogonal arms lengths contribute independent uncorrelated
parts to $\Xi(\tau)$. We assume that the transverse variances from
Eq. (\ref{eq:norm}) directly apply to the reference world line, as
light beams make simultaneous measurements of orthogonal directions
at this position.
\begin{equation}
\xi_{0}\leq2\ell_{P}\equiv ct_{P}/\sqrt{\pi}
\end{equation}
Note that we may not count the arm length $L$ twice to account for
the round trip time, because doing so while also counting two orthogonal
arms would violate quantum limits on information. The holographic
bound limits us to the size of the apparatus instead of the whole
light path. This normalization is consistent with previous work \cite{KwonHogan2014}.

The lower bound is saturated when the transverse variance from Eq.
(\ref{eq:norm}) is divided between the reference world line (e.g.
the beamsplitter) and the end mirror, implying that the equivalent
measurable position uncertainty at the reference world line is halved.
We also assume that holographic bounds on information limit the measurable
noise to the transverse variance from the arm length in one direction,
seeing the two orthogonal directions as noncommutative.
\begin{equation}
\xi_{0}\geq\frac{1}{2}\ell_{P}\equiv ct_{P}/\sqrt{16\pi}
\end{equation}
We consider this the smallest normalization we can reasonably expect
(even under the most conservative assumptions) when all of the holographic
uncertainty is manifest in shear mode correlations. If the measured
noise is below this level, it likely indicates that at least some
of the uncertainty has been squeezed into other unmeasured degrees
of freedom such as rotational modes.

The Holometer is expected to reach the sensitivity to detect or rule
out all normalizations that are physically plausible.

\subsection{General Class: Models Spanning $\pm2L$ }

In this general class of models, the causal support contains all events
that influence the signal at a given time, and therefore the correlation
does not necessarily drop to zero at the edge of a single causal diamond
(representing a single round trip to end mirrors). Instead, we consider
the overlap between causal diamonds separated by $\tau$, and maintain
functional support for $\Xi(\tau)$ wherever the shared 4-volume is
nonzero. We expect $\Xi(\tau)$ to be continuous at the edge of a
single causal diamond (since the amount of overlap is continuous as
a function of $\tau$), and linearly fall off to zero at the edge
of the causal limit.

In previous work\cite{KwonHogan2014}, we presented as an illustrative
model a simple triangle spanning $\pm2L$, physically interpreted
as a new kind of exotic ``jitter'' in space and in reference world
line position. This maintains the same form of straightforward linearity
used in the previous section, just with longer functional support.
In particular, the maximum normalization of this model, corresponding
to $\Xi(\tau=0)/L=2\ell_{P}\equiv ct_{P}/\sqrt{\pi}$, was used as
a ``baseline model'' to estimate experimental sensitivity for design
purposes and tested during the initial run of the Holometer\cite{ChouEtal2015,ChouEtal:2016}.

\begin{figure}
\includegraphics[scale=0.5]{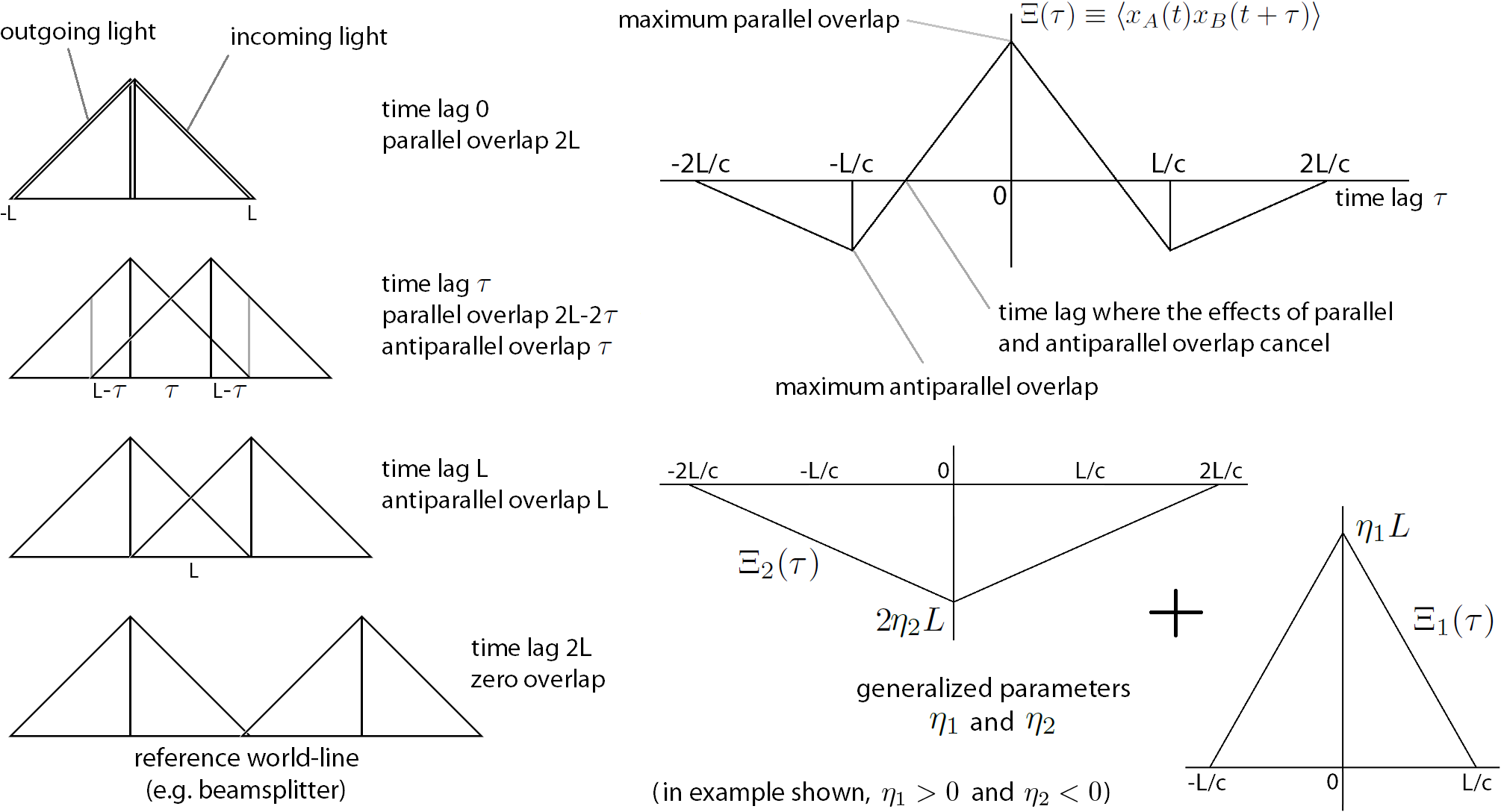}

\protect\caption{A generalized class of models, spanning $\pm2L$: Creating a two-parameter
space for all possible correlation spectra. The ``baseline model''
presented in \cite{KwonHogan2014} and tested in \cite{ChouEtal2015},
a case in which parallel and antiparallel overlaps give equal (positive)
contributions, corresponds to a $\pm2L/c$ triangle with $\eta_{1}=0$,
$\eta_{2}=+\ell_{P}\equiv ct_{P}/\sqrt{4\pi}$.\label{fig:timedomain-2l/c}}
\end{figure}

Figure \ref{fig:timedomain-2l/c} shows the full range of models reflecting
a generalized understanding of linearity in this framework of overlapping
causal diamonds (and consistent with the constraints outlined in Section
\ref{sub:Constraints}). The sequence of diagrams on the left hand
side shows how the overlap between two causal diamonds changes as
the time lag between them increases. In each causal diamond describing
an interferometer, the left half represents an outgoing tracer photon
propagating towards the end mirrors, and the right half represents
an incoming tracer photon coming back towards the reference world
line. We use the term ``parallel overlap'' and ``antiparallel overlap''
to respectively denote ranges of time where the light path is parallel
or antiparallel in direction while the tracer photons in the two interferometers
traverse an overlapping 4-volume of spacetime.

An important illustrative case is when the parallel and antiparallel
overlaps give contributions to $\Xi(\tau)$ that are opposite in sign
(if they are equally weighted, they precisely cancel each other at
time lag $\tau=\frac{2}{3}L/c$). The top plot on the right hand side
shows an example of $\Xi(\tau)$ under such assumptions. It is naturally
a sum of the two plots below it, which define the generalized slope
parameters $c\eta_{1}$ and $c\eta_{2}$ that establish the parameter
space for this class of models. As before, both parameters are on
the order of Planck time and not dependent on the apparatus size,
and we do not \textit{a priori} constrain them to positive or negative
values (despite the illustrative example). A general class of models that 
contains a linear combination of $\pm L$ and $\pm 2L$ behavior can be 
written as a parameterized sum,
\begin{eqnarray}
\Xi(\tau) & = & \Xi_{1}(\tau)+\Xi_{2}(\tau) \nonumber \\
 & = & \begin{cases}
\eta_{1}(L-c\left|\tau\right|) & \left|\tau\right|<\frac{L}{c}\\
0 & \left|\tau\right|>\frac{L}{c}
\end{cases}+\begin{cases}
\eta_{2}(2L-c\left|\tau\right|) & \left|\tau\right|<\frac{2L}{c}\\
0 & \left|\tau\right|>\frac{2L}{c}
\end{cases}
\end{eqnarray}
\begin{equation}
c\eta_{1}\equiv\dot{\Xi}_{1}(-\frac{L}{c}<\tau<0)=-\dot{\Xi}_{1}(0<\tau<\frac{L}{c})\qquad\textrm{and}\qquad c\eta_{2}\equiv\dot{\Xi}_{2}(-\frac{2L}{c}<\tau<0)=-\dot{\Xi}_{2}(0<\tau<\frac{2L}{c})
\end{equation}

The two triangle functions, with causal support ranging to $\pm L/c$
and $\pm2L/c$ respectively, each satisfy the condition for scaling
symmetry established in the previous section. We interpret this to
mean that we are allowing for two ways in which nonlocal correlations
can manifest in the emergent geometry, one that is fully contained
within the causal 4-volume of a single diamond (a single round trip
to end mirrors) and another that has a causal memory of $\tau=\pm2L/c$
and depends on the overlap between two causal diamonds.

The former of these effects, reflecting the value of $\eta_{1}$ (or
the subspace in which $\eta_{2}=0$), was treated in the previous
section on simple models spanning $\pm L/c$. Such cases were discussed
in previous work\cite{KwonHogan2014} as representing possibilities
in which each reflection (when the light interacts with an observer
or mass at the end mirror or at the reference world line) sets a boundary
condition that resets all quantum geometric information, thereby zeroing
the device's response to antiparallel overlap between tracer photon
paths before and after an end mirror reflection.

We can now extend this one-dimensional (sub)space into a generalized
two-dimensional space defined by universal parameters $\eta_{1}$
and $\eta_{2}$. Similar to the previous class of models, a choice
of universal parameter values predicts a frequency spectrum that can
be compared with Holometer data. Here, the predictions and data can
be used to generate joint likelihood contours.

\begin{figure}
\begin{centering}
\includegraphics[scale=0.36]{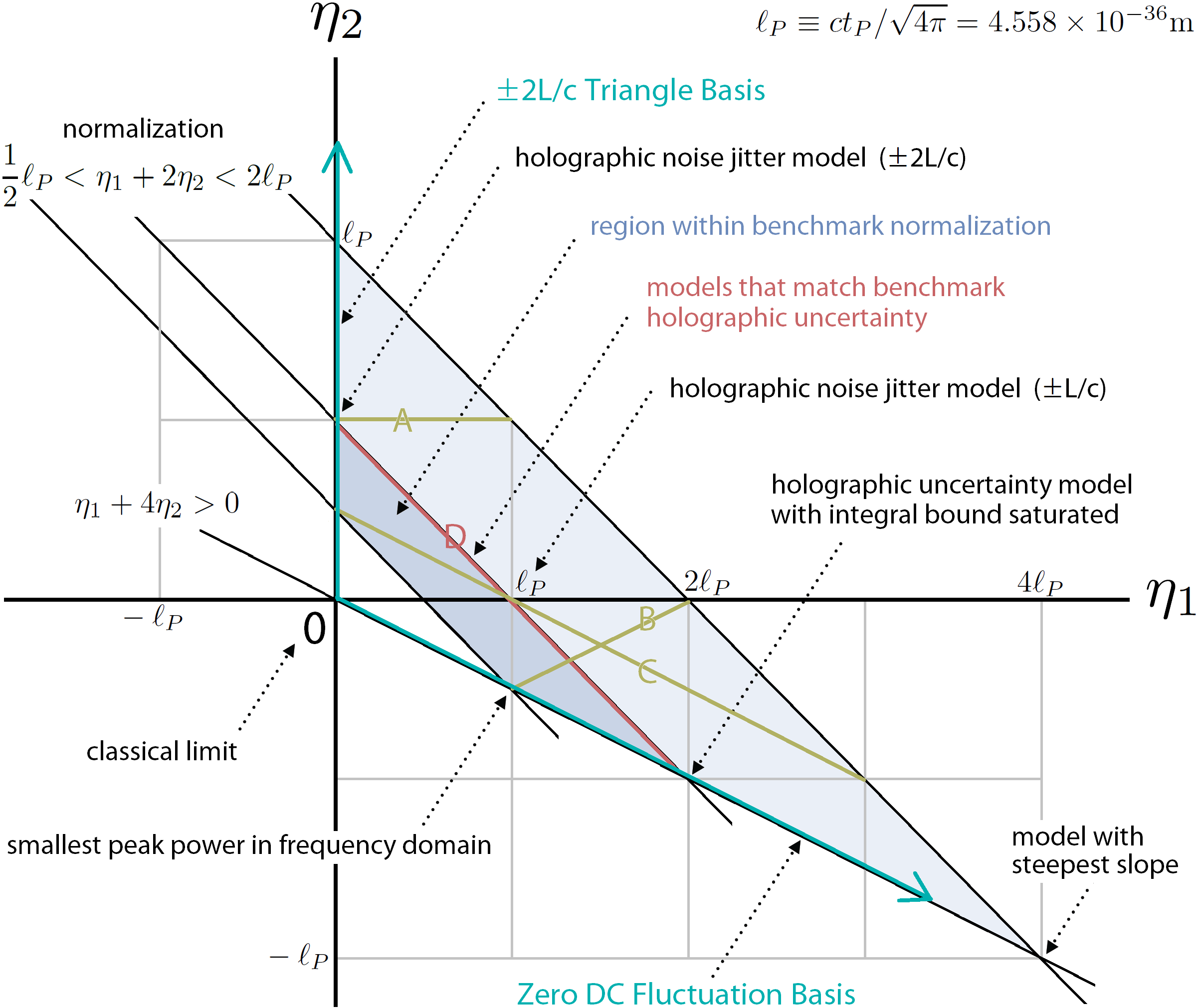} 
\par\end{centering}

\protect\caption{General parameter space: Two observable length parameters ($\eta_{1}$
and $\eta_{2}$) span all possible models of noise that are consistent
with symmetries of spacetime, causal structure, and holographic limits
on geometric information. The viable parameter space is defined by
two basis vectors (the ``$\pm2L$ triangle'' and ``zero-DC'' models)
and and upper and lower ranges on normalization. Sample spectra along
the four labeled lines are given in Figure \ref{fig:freqdomain-2l/c}.\label{fig:paramplot-2l/c}}
\end{figure}

\begin{figure}
\begin{centering}
\includegraphics[scale=0.4]{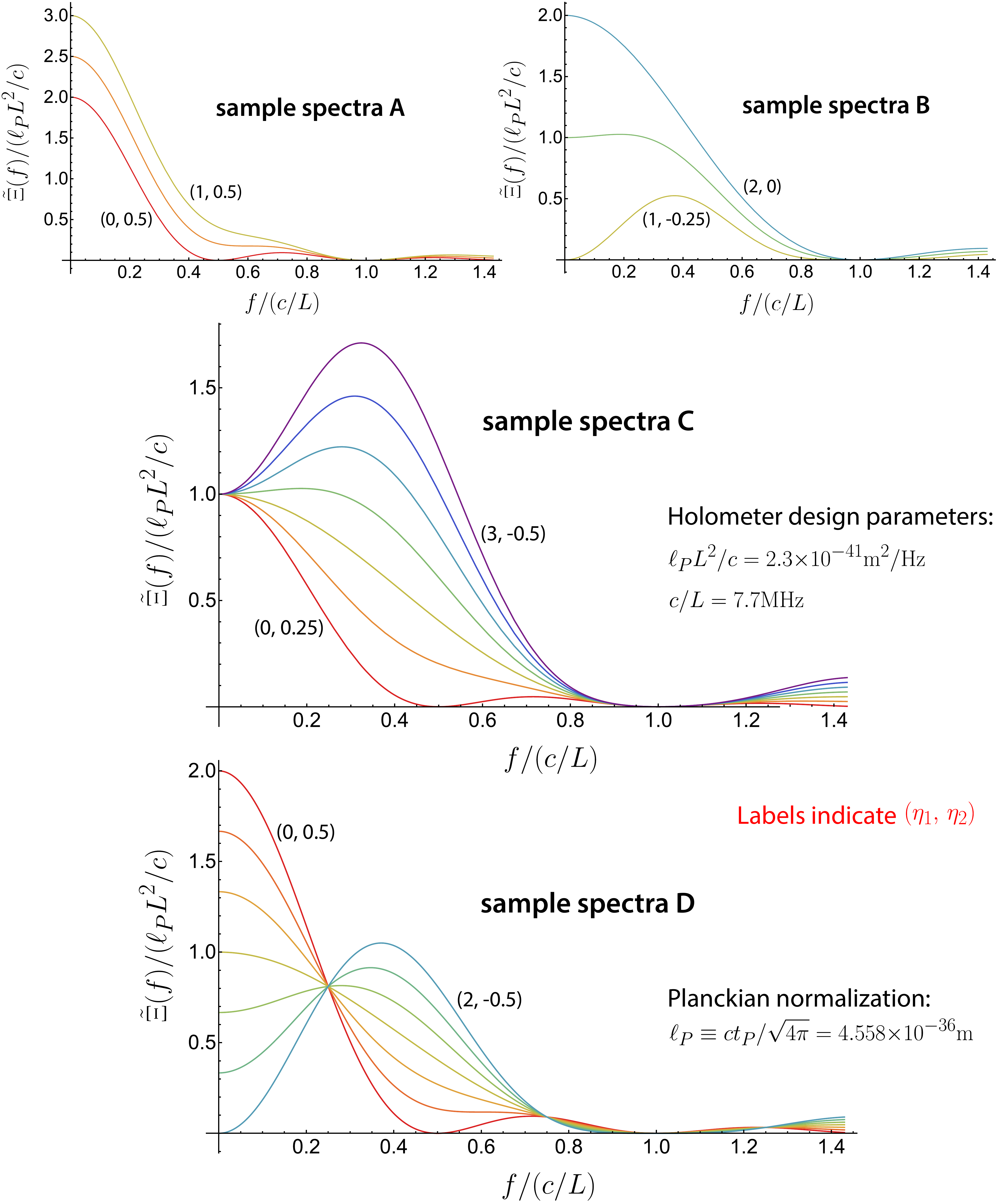} 
\par\end{centering}

\protect\caption{A representative sample of possible frequency domain spectra. Values
of $\eta_{1}$ and $\eta_{2}$ are chosen along the four labeled lines
within the parameter space in Figure \ref{fig:paramplot-2l/c}, linearly
from left to right, and color-coded along the visible spectrum based
on the horizontal location of the model (comparing plots of the same
color illustrates differences according to vertical location). Vertical
axes are scaled relative to the variance per unit frequency corresponding
to $\ell_{P}\equiv ct_{P}/\sqrt{4\pi}$ and system scale $L$. Horizontal
axes are scaled to inverse light crossing time for $L$.\label{fig:freqdomain-2l/c}}
\end{figure}

The constraints on the parameter space derive from similar considerations
as before, and the allowed region is plotted in Figure \ref{fig:paramplot-2l/c}.
The integral $\int d\tau\Xi$ must be nonnegative, which implies:
\begin{equation}
\eta_{1}+4\eta_{2}\geq0
\end{equation}
When the bound is saturated ($\eta_{1}+4\eta_{2}=0$), there is zero
DC power in the frequency spectrum, because the integral vanishes.
This corresponds to the illustrative case above, where the parallel
and antiparallel overlaps give equal-weight contributions of opposite
sign. We physically interpret this as a kind of exotic ``jitter''
in space (and in world line positions) similar to models considered
in previous work\cite{KwonHogan2014}, but with a different device
coupling that flips sign as the light beam reverses direction after
an end mirror reflection.

The transform $\tilde{\Xi}(f)$ must be real, even, and nonnegative
at all frequencies for $\Xi(\tau)$ to be an autocorrelation of a
real-valued wide-sense stationary random process \cite{OppenheimVerghese2015}.
This requires:
\begin{equation}
\eta_{1}\geq0\qquad\textrm{and}\qquad\eta_{1}+4\eta_{2}\geq0
\end{equation}
The first part of this constraint makes intuitive sense--- we expect
stronger positive correlations at time lag under $L/c$, when there
is parallel overlap. When this non-redundant segment of the bound
is saturated ($\eta_{1}=0$, $\eta_{2}>0$), the function reduces
to the illustrative $\pm2L$ triangle models considered in previous
work, including the ``baseline model'' \cite{KwonHogan2014}. These
models get equal positive contributions from parallel and antiparallel
overlaps, implying a constant positive device coupling throughout.

The analytic form of the frequency domain power spectrum is helpful
for demonstrating these features: 
\begin{eqnarray}
\tilde{\Xi}(f) & = & 2\int_{0}^{\infty}(\Xi_{1}(\tau)+\Xi_{2}(\tau))\cos(2\pi f\tau)\textrm{d}\tau \nonumber \\
 & = & \frac{2c\eta_{1}}{(2\pi f)^{2}}\left[1-\cos\left(\frac{f}{c/2\pi L}\right)\right]+\frac{2c\eta_{2}}{(2\pi f)^{2}}\left[1-\cos\left(\frac{f}{c/4\pi L}\right)\right] \nonumber \\
 & = & c\eta_{1}\left(\frac{L}{c}\right)^{2}\textrm{sinc}^{2}\left(\frac{f}{c/\pi L}\right)+c\eta_{2}\left(\frac{2L}{c}\right)^{2}\textrm{sinc}^{2}\left(\frac{f}{c/2\pi L}\right)
\end{eqnarray}

Figure \ref{fig:freqdomain-2l/c} shows a sampling of the type of
spectra that are expected. The four plots are a representative sample
of the mathematical characteristics featured in $\tilde{\Xi}(f)$
as $\eta_{1}$ and $\eta_{2}$ span the range of allowed parameter
space.

\subsubsection*{Basis Vectors for the General Parameter Space}

\begin{figure}
\includegraphics[scale=0.37]{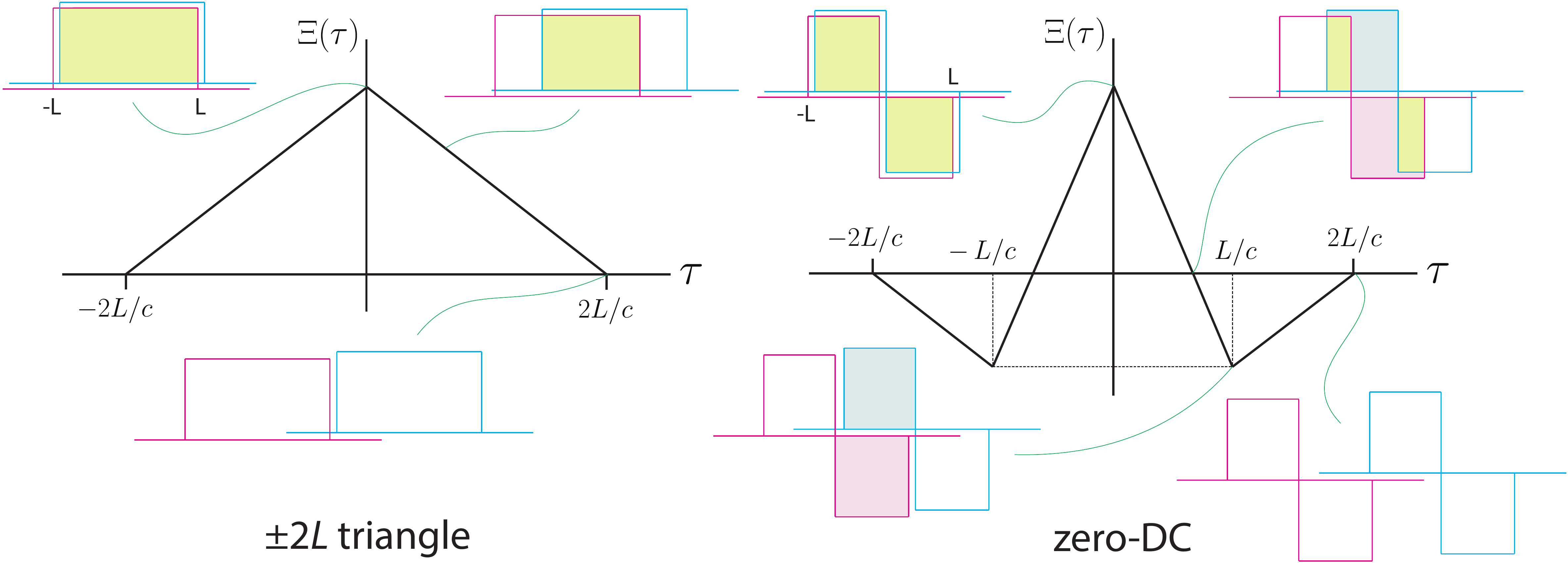}

\protect\caption{The basis vectors: The ``zero-DC'' and ``$\pm2L$ triangle'' models,
corresponding to flat accumulations of exotic jitter, are generated
from autocovariances of $\pm L$ boxcar functions that differ by a
sign change at the midpoint.\label{fig:basis-vectors}}

\end{figure}

The two bounding cases--- the ``zero-DC model'' ($\eta_{1}+4\eta_{2}=0$,
$\eta_{1}>0$, $\eta_{2}<0$) and the ``$\pm2L$ triangle model''
($\eta_{1}=0$, $\eta_{2}>0$)--- can be considered basis vectors
for this general parameter space. Both correspond to a wide-sense
stationary stochastic accumulation of spacetime jitter with constant-magnitude
device coupling throughout the light path (or, nonlocally, over measurement
time), the only difference being whether the coupling undergoes a
sign change halfway through the light storage time (at the end mirror
reflection). Figure \ref{fig:basis-vectors} demonstrates this point.
The constant accumulation of fluctuations throughout each round trip
is represented by flat boxcar functions of $\pm L$ support (with
and without the sign change at the midpoint), and the two basis models
(both spanning $\pm2L$) are generated from their autocovariances.

These basis vectors can be considered ``natural'' models where the
covariance of exotic correlations follow the causal structure of the
spacetime being measured. Here, we use ``covariance'' in a statistical
sense, as opposed to the general relativistic meaning used throughout
the rest of this paper. Adopting the perspective that spacetime is
``relational''\cite{Rovelli2004,Thiemann2008,Ashtekar2012}, a quantum
system woven out of Planckian elements, the spatial structure of their
entanglement can be expressed as a statistical covariance among observable
operators, with a coherence length and time of the Planck scale. This
bottom-up construction from Planckian subsystems, with covariance
structures defined by causal surfaces, will be useful in upcoming
work as we generalize our predictions to rotational correlations\cite{Hogan2015,HoganKwonRichardson2016}.

\subsubsection*{Normalization for Correlated Interferometer Noise}

As with the previous simple class of models (spanning $\pm L$), we
will use Eq. (\ref{eq:norm}) and a similar set of assumptions to
establish benchmark normalizations and upper and lower ranges for
$\eta_{1}$ and $\eta_{2}$ in the general two-parameter space (spanning
$\pm2L$). The viable region of parameter space is plotted in Figure
\ref{fig:paramplot-2l/c}. 

We first note that, similar to the simpler class, 
\begin{equation}
\frac{1}{2}\ell_{P}\leq\Xi(\tau=0)/L=\eta_{1}+2\eta_{2}\leq2\ell_{P}\quad\;\mbox{where}\quad\;\ell_{P}\equiv ct_{P}/\sqrt{4\pi}.
\end{equation}
This, in addition to the constraints from statistical requirements
above, automatically sets appropriate ranges for the individual parameters:
$0\leq\eta_{1}\leq4\ell_{P}$ and $-\ell_{P}\leq\eta_{2}\leq\ell_{P}$.
The upper estimate of $\eta_{2}=\ell_{P}\equiv ct_{P}/\sqrt{4\pi}$
(with $\eta_{1}=0$) is consistent with the ``baseline model'' in
previous work\cite{KwonHogan2014} and was tested during the initial
run of the Holometer\cite{ChouEtal2015}. 

As before, as our benchmark normalization matching holographic position
uncertainty, we use a straightforward comparison between Eq. (\ref{eq:norm})
and $\Xi(\tau=0)/L$: 
\begin{equation}
\eta_{1}+2\eta_{2}=\ell_{P}\equiv ct_{P}/\sqrt{4\pi}\:,\qquad\left|\Xi(\tau)\right|\leq\ell_{P}L
\end{equation}

The Holometer is expected to reach the sensitivity to confirm or reject
all physically viable models, exhaustively probing the shear degree
of freedom all the way down to our lower bound normalization before
leading into the next stage of the experimental program that will
test rotational correlations. Quantum geometrical uncertainties can
be squeezed into certain degrees of freedom or others due to underlying
symmetries in nature or limitations in device coupling; we seek to
empirically identify the ones where the information is bounded, guiding
the theoretical development.

\section{Current Limits and Holometer Sensitivity}

The measured spectra at the Holometer are plotted in units of differential
arm length (DARM) variance per unit frequency, following the standard
conventions adopted by the interferometry community \cite{ChouEtal2015,ChouEtal:2016}.
Since the observables considered in the above predictions for $\Xi(\tau)$
are antisymmetric port signals calibrated to length units, we need
to make the following conversion from $\tilde{\Xi}(f)$ to measured
power spectral density (PSD):
\begin{equation}
\textrm{PSD}(f)=\tilde{\Xi}(f)\cdot2\cdot\left(\frac{1}{2}\right)^{2}=\frac{1}{2}\tilde{\Xi}(f)
\end{equation}
This is because the dark port signals reflect the optical path difference
(OPD) over the full round trip light paths in the two arms. The standard
phase-to-length transfer function for Michelson interferometers assumes
low-frequency response, in which case $\textrm{OPD}=2\cdot\mbox{DARM}$.
Since the measured PSD and $\tilde{\Xi}(f)$ are in units of power
(or variance), this factor of $\frac{1}{2}$ gets squared. The remaining
additional factor of 2 comes from the fact that $\tilde{\Xi}(f)$
is a standard Fourier transform defined for both positive and negative
frequencies, whereas the measured PSD is expressed in the engineering
convention, defined only for positive frequencies after folding into
the domain the power contained in the redundant negative frequencies.

\begin{figure}[p]
\includegraphics[scale=0.9]{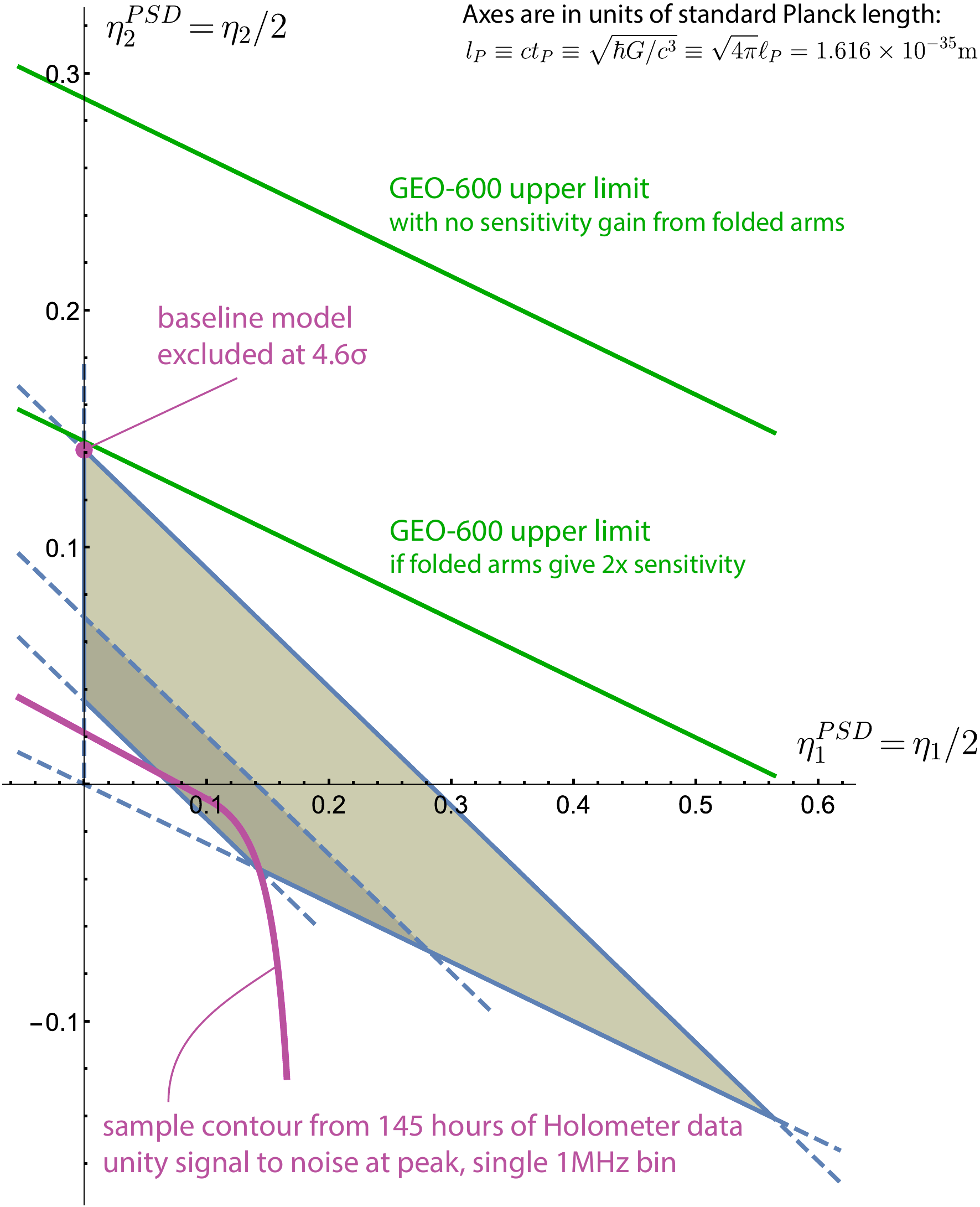}

\protect\caption{Current limits and Holometer sensitivity: Upper limits on holographic
noise from GEO-600\cite{FrolovGrote2014} are given under specific
assumptions about whether the folded arms lead to a twofold increase
in signal power. The illustrative example of ``baseline model''
from \cite{KwonHogan2014} was tested in \cite{ChouEtal2015} and
ruled out at $4.6\sigma$. Also shown is an example contour of Holometer
sensitivity after 145 hours of integration, assuming unity signal-to-noise
ratio in a single 1MHz bin at the peak (accurate to $\pm10\%$) \cite{ChouEtal2015,ChouEtal:2016}.
This reflects what we would observe in the case of a null result,
as longer integration times are expected to move the contour closer
towards zero. If there is a significant detection, the actual likelihood
contour would converge around a single point in the parameter space,
possibly much farther away from the origin than the plotted example.
A highly significant verification or exclusion of the holographic
shear noise hypothesis is possible within realistic integration times.
Axes are in units of standard Planck length, scaled to power spectral
density (PSD) values measurable at the Holometer.\label{fig:explimits}}
\end{figure}

Figure \ref{fig:explimits} plots the parameter space constructed
above in terms of measured PSD, using universal Planck length units.
It demonstrates how this parameter space is used to interpret experimental
data and draw constraints on different phenomenological models. 

The two lines labeled ``GEO-600 noise'' reflect upper limits from
\cite{FrolovGrote2014}. Prior to the Fermilab Holometer, GEO-600,
employing a single Michelson interferometer with folded arms, was
the experiment most sensitive to this type of Planckian geometrical
correlation \cite{KwonHogan2014}. While it has proven difficult to
comprehensively identify and quantify sources of other environmental
and technical noise\cite{Hild2009}, the collaboration has produced
a rough upper limit on unidentified flat-spectrum noise (the ``signal'')
around the $1\sim6\textrm{kHz}$ range where shot noise is dominant
and can be reliably modeled, at $1.25\times10^{-22}\textrm{Hz}^{-1/2}$\cite{FrolovGrote2014}
(in strain units). Because this upper limit is only available in a
relatively narrow band at a frequency much lower than $c/L$ (around
0.5MHz for GEO-600), it can only specify one point on the frequency
spectrum and corresponds to a straight line in the parameter space.
We plot two lines because it is not clear whether the folded arms
in GEO-600 would contribute to a twofold increase in the variance
measured in differential arm length. Such considerations were discussed
in previous work \cite{KwonHogan2014}.

The point labeled ``baseline model'' refers to the illustrative
example discussed in previous work\cite{KwonHogan2014} and above.
It was tested as the nominal model for the initial run of the Holometer.
According to the recently released first results, this particular
model was excluded at $4.6\sigma$, based on 145 hours of data \cite{ChouEtal2015,ChouEtal:2016}.

For the Holometer, we have also plotted an example contour based on
the noise spectrum released with the initial results\cite{ChouEtal2015,ChouEtal:2016},
assuming unity signal-to-noise at the peak of the predicted spectrum
corresponding to each point on the contour. Because this sensitivity
example is available across a much wider range of frequencies ($1\sim6\textrm{MHz}$),
we are able to generate contours that cover much wider classes of
models that reach maximal power at different frequencies (e.g. Figure
\ref{fig:freqdomain-2l/c}) instead of a straight line in parameter
space like the GEO-600 example.

This is not an exclusion plot, or a likelihood contour at any significance
level, and it does not even reflect shot noise-subtracted residual
noise like the GEO-600 lines. It is merely intended to demonstrate
noise levels attainable at the Holometer, and shows that it takes
145 hours of integration time to get a roughly $1\sigma$ result in
a 1MHz-wide band at the peak of each modeled spectrum along this contour.

%This data set certainly does not have sufficient sensitivity to probe the smaller normalizations within the parameter space, and should be considered accurate to about $\pm10\%$ depending on the exact calibration and systematic budget.
 %A rigorous Bayesian analysis that optimally weighs all frequency bins within the $1\sim11\textrm{MHz}$ data band along the modeled spectra, representing 700 hours of integration time, is expected in upcoming results.

The Holometer operates almost entirely within the frequency range
dominated by shot noise. Using the behavior of cross-correlated and
integrated shot noise described in \cite{ChouWeiss2009,Kamai2013,KwonHogan2014,ChouEtal2015,ChouEtal:2016},
it is straightforward to extrapolate the noise levels demonstrated
in this example to longer integration times and different choices
for frequency binning. An \textit{N}-fold increase in integration
time (or frequency bin width) generally reduces the uncorrelated shot
noise power (or error bars in the binned power spectrum) by a factor
of $1/\sqrt{N}$, while the correlated exotic Planckian noise remains
the same. If we assume that geometry behaves classically up to the
level of sensitivity we can project from such extrapolations, a realistic data set would be sufficient to rule out the smallest normalizations
within the viable parameter space at roughly $3\sigma$, with the
likelihood contour approaching the classical limit at the origin.
On the other hand, if a larger signal is detected, the data-based
likelihood contour would likely never even reach the example plotted,
and instead converge around a specific point within the viable parameter
space, with the statistical significance expected to be at least comparable
but possibly larger depending on the exact location (a 2000-hour data
set would be sufficient for $5\sigma$ detection at the edges of the
parameter space closest to the origin). The example contour thus demonstrates
that the Holometer has the capacity to probe all physically plausible
regions within the ``shear mode'' parameter space, within reasonable
integration times.

\section{Future Configuration: Rotational Modes}

\begin{figure}
\begin{centering}
\includegraphics[scale=0.29]{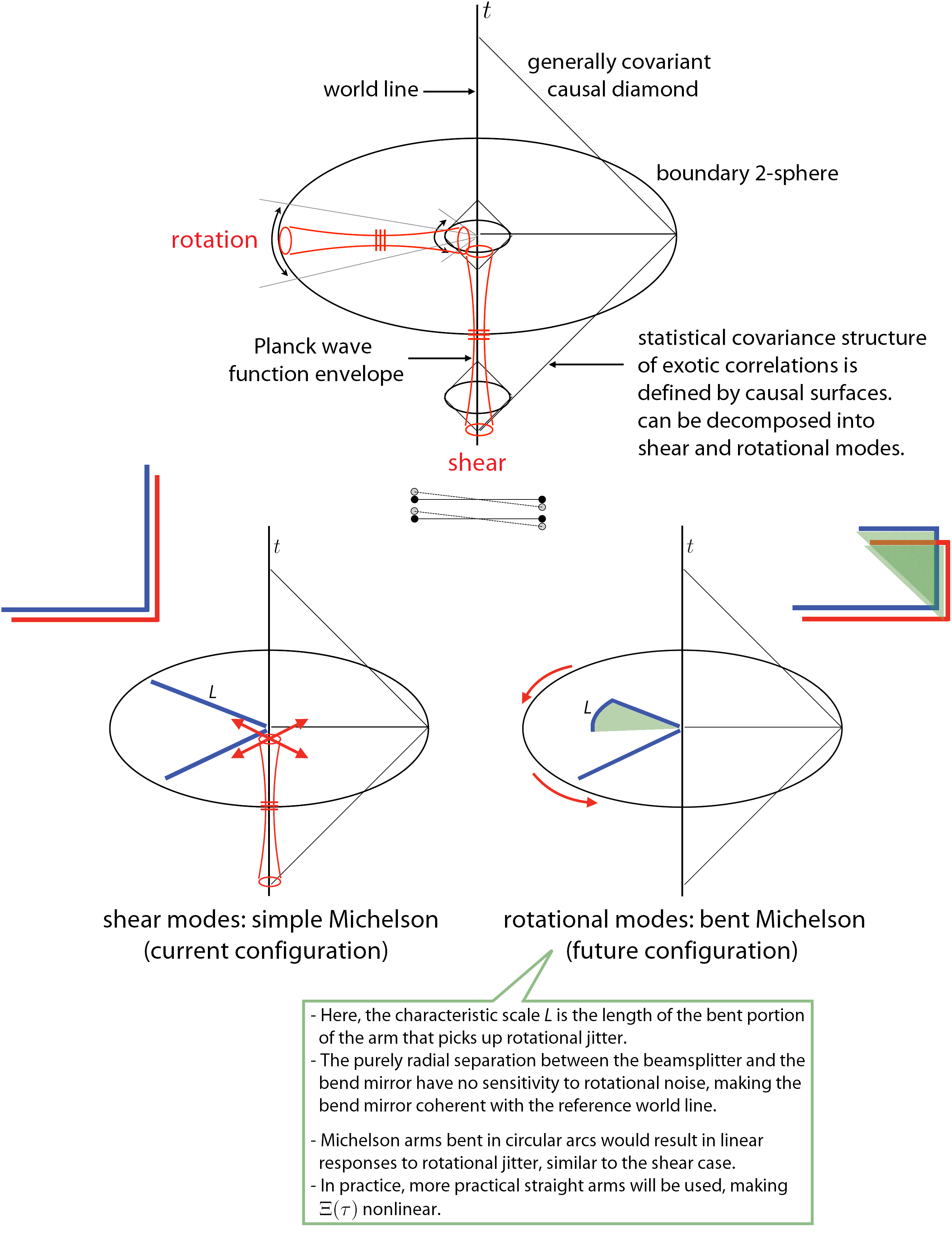}
\par\end{centering}

\protect\caption{The statistical covariance structures of exotic Planck-scale geometrical
correlations, following null surfaces of causal relations, is shown
to decompose into orthogonal modes of shear and rotation. Interferometer
setups that couple to each correlation mode are shown: the simple
Michelson and bent Michelson setups are capable of detecting shear
and rotational noise respectively. Also shown is a spatially extended
Sagnac interferometer, optimal for probing Planckian departures from
perfect symmetry of inertia.\label{fig:future-configurations}}
\end{figure}

Figure \ref{fig:future-configurations} depicts a conceptual big-picture
understanding of the overall hypothesis--- that geometrical information
about directions around world lines is constrained by Planckian bounds
defined via general relativistically covariant causal diamonds. In
a quantum system of ``relational'' geometry woven out of causal
diamonds containing Planckian subsystems, their spatial entanglement
follows statistical covariance structures defined by causal surfaces
(note again the two different senses in which ``covariance'' is
used, unlike the general relativistic usage elsewhere in this paper).
These null relationships can be decomposed into two orthogonal correlation
modes: shear and rotation.

Shear modes couple \textit{transversely} to the arms of a simple Michelson,
where light propagates purely radially within the causal diamond.
This transverse jitter is manifest as a Planckian spreading of the
reference world line, where orthogonal light beams meet at the beamsplitter.
The $180\textdegree$ reflections at the end mirrors do not capture
the transverse uncertainty, naturally.

Rotational modes\cite{Hogan2015,HoganKwonRichardson2016}, on the other hand, cannot be seen
in a simple Michelson, because neither the beamsplitter nor the end
mirrors have any sensitivity to this type of correlations of quantum-geometrical
jitter. To detect them, we add a $90\textdegree$ reflection \textit{away}
from the beamsplitter, at significant radial separation, and have
a significant portion of the light path involve \textit{angular} propagation.
The characteristic scale $L$ for the accumulation of jitter would
be the distance between the end mirror and the \textit{bend mirror},
as the latter would be coherent with the reference world line for
this type of quantum-geometrical uncertainty. The directional uncertainty,
defined in the inertial frame of the reference world line, in this
case couples \textit{longitudinally} to the bent portion of the arm.
We interpret this as an accumulated effect of the quantum-geometric
``twists'' of the Planckian elements that comprise the area circumscribed
by the bent arm.

In an idealized device, the bent portion of the arm would follow a
circular arc, as shown inside a causal diamond in Figure \ref{fig:future-configurations}.
This would ensure a linear form of $\Xi(\tau)$, as the requirements
laid out above using the scaling symmetry of causal diamonds would
naturally extend here. In fact, if it were possible to have such a
device without constant interaction between light and matter, the predicted
spectrum would neatly fall into a parameter space spanned by the basis
vectors we constructed for the shear case (albeit with different normalizations;
see caveats on shear mode normalization above). In fact, we expect
a response similar to the ``zero-DC model,'' as will be detailed
in upcoming work.
The precise shape of the predicted spectrum in realistic bent arm configurations (with straight arms) will be calculated
in upcoming work.

%The bent Michelson configuration is not the most optimal setup for
%detecting nonlocal rotational correlations. In the case of a detection,
%we propose a second reconfiguration to a spatially extended Sagnac
%setup, traditionally used for measurements of light in non-inertial
%frames. This configuration samples the background spacetime differently
%in one important way: instead of measuring phase jitters accumulated
%over a round trip made by one light beam, it compares two split light
%beams going around a common enclosed area in opposite directions before
%recombining at the beamsplitter. \medskip{}

The Holometer experimental program is designed to complete a comprehensive
study of the hypothesized exotic correlations. As seen above, general
arguments can set the relevant scaling and amplitude for holographic
noise, but not the spatial pattern. There are limited choices for
viable spatial patterns: strain, shear, and rotation. 

The strain case is not a part of our Planckian \textit{directional}
uncertainty hypothesis, as a coherent and directionally isotropic
strain uncertainty likely violates general covariance, but it was
nevertheless explored with existing data in previous work \cite{KwonHogan2014}.
Strain noise that scales like $\langle\Delta x^{2}\rangle\sim ct_{P}L$,
like we consider here, was ruled out by data from the initial LIGO
detectors. Because longitudinal strain noise does not dimensionally
reduce the total degrees of freedom, there are arguments for a different
scaling behavior $\langle\Delta x^{2}\rangle\sim(c^{2}t_{P}^{2}L)^{2/3}$
based on a more simplistic interpretation of holographic bounds\cite{NgDam2000,Ng2002,KwonHogan2014},
but those models do not retain coherence, predicting locally observable
effects instead of a macroscopic form of nonlocality. These predicted
violations of locality have been experimentally ruled out \cite{LieuHillman2003,RagazzoniTurattoGaessler2003,NgChristiansenDam2003,ChristiansenNgDam2006,ChristiansenNgFloydPerlman2011,PerlmanNgFloydChristiansen2011,PerlmanRappaportChristiansenNg2015}.

The upcoming results from a long stretch of data under the current simple
Michelson design, together with the planned reconfiguration to a bent Michelson
setup, should comprehensively probe the two more plausible modes of
correlation: shear and rotation. They will either detect specific
statistical covariance structures of Planckian correlations in spacetime,
or provide concrete evidence for near-perfect Planck scale symmetries
of causality and inertia.
\begin{acknowledgments}
We are grateful to Aaron Chou, Richard Gustafson, Brittany Kamai,
Robert Lanza, Stephan Meyer, Lee McCuller, Jonathan Richardson, Chris
Stoughton, Raymond Tomlin, Samuel Waldman, Rainer Weiss, and other
contributors to the Holometer program, especially for their detailed
inputs on the various aspects of interferometer design, noise behavior,
and signal interpretation. We thank Hartmut Grote for his explanations
regarding GEO-600 noise models and data interpretation. This work
was supported by the Department of Energy at Fermilab under Contract
No. DE-AC02-07CH11359, and at the University of Chicago by grant No.
51742 from the John Templeton Foundation. O.K. is supported by the
Basic Science Research Program (Grant No. NRF-2016R1D1A1B03934333) 
of the National Research Foundation of Korea (NRF) funded by the Ministry of Education.
\end{acknowledgments}

\bibliographystyle{unsrturl}
\bibliography{bib_o_kwon}

\begin{thebibliography}{10}

\bibitem{Rovelli2004}
C.~Rovelli.
\newblock {\em Quantum Gravity}.
\newblock Cambridge University Press, 2004.

\bibitem{Thiemann2008}
T.~Thiemann.
\newblock {\em Modern Canonical Quantum General Relativity}.
\newblock Cambridge University Press, 2008.

\bibitem{Ashtekar2012}
A.~Ashtekar.
\newblock {Introduction to Loop Quantum Gravity}.
\newblock {\em PoS}, QGQGS2011:001, 2011.
\newblock arXiv:1201.4598 [gr-qc].

\bibitem{DeservanNieuwenhuizen1974}
S.~Deser and P.~van Nieuwenhuizen.
\newblock Nonrenormalizability of the quantized {D}irac-{E}instein system.
\newblock {\em Phys. Rev. D}, 10:411, 1974.

\bibitem{Weinberg1996}
S.~Weinberg.
\newblock {\em The Quantum Theory of Fields}.
\newblock Cambridge University Press, 1996.

\bibitem{Wilczek1999}
F.~Wilczek.
\newblock Quantum field theory.
\newblock {\em Rev. Mod. Phys.}, 71:S85, 1999.

\bibitem{EllisMavromatosNanopoulos1992}
J.~Ellis, N.~Mavromatos, and D.~V. Nanopoulos.
\newblock String theory modifies quantum mechanics.
\newblock {\em Phys. Lett. B}, 293:37, 1992.

\bibitem{Polchinski1998}
J.~Polchinski.
\newblock {\em String Theory}.
\newblock Cambridge University Press, 1998.

\bibitem{Hossenfelder2013}
S.~Hossenfelder.
\newblock Minimal length scale scenarios for quantum gravity.
\newblock {\em Living Rev. Rel.}, 16:2, 2013.

\bibitem{DouglasNekrasov2001}
M.~R. {Douglas} and N.~A. {Nekrasov}.
\newblock Noncommutative field theory.
\newblock {\em Rev. Mod. Phys.}, 73:977, 2001.

\bibitem{GiddingsMarolfHartle2006}
S.~B. Giddings, D.~Marolf, and J.~B. Hartle.
\newblock Observables in effective gravity.
\newblock {\em Phys. Rev. D}, 74:064018, 2006.

\bibitem{Giddings2007}
S.~B. {Giddings}.
\newblock Black holes, information, and locality.
\newblock {\em Mod. Phys. Lett. A}, 22:2949, 2007.

\bibitem{Banks2009}
T.~Banks.
\newblock Deriving particle physics from quantum gravity: A plan.
\newblock arXiv:0909.3223 [hep-th], 2009.

\bibitem{Banks2011}
T.~Banks.
\newblock Holographic space-time: The takeaway.
\newblock arXiv:1109.2435 [hep-th], 2011.

\bibitem{BardeenCarterHawking1973}
J.~M. Bardeen, B.~Carter, and S.~Hawking.
\newblock The four laws of black hole mechanics.
\newblock {\em Commun. Math. Phys.}, 31:161, 1973.

\bibitem{Bekenstein1973}
J.~D. Bekenstein.
\newblock Black holes and entropy.
\newblock {\em Phys. Rev. D}, 7:2333, 1973.

\bibitem{Bekenstein1974}
J.~D. Bekenstein.
\newblock Generalized second law of thermodynamics in black-hole physics.
\newblock {\em Phys. Rev. D}, 9:3292, 1974.

\bibitem{Hawking1974}
S.~Hawking.
\newblock Black hole explosions.
\newblock {\em Nature}, 248:30, 1974.

\bibitem{Hawking1975}
S.~Hawking.
\newblock Particle creation by black holes.
\newblock {\em Commun. Math. Phys.}, 43:199, 1975.

\bibitem{tHooft1993}
G~'t~Hooft.
\newblock Dimensional reduction in quantum gravity.
\newblock In {\em Conference on Particle and Condensed Matter Physics
  (Salamfest)}, 1993.
\newblock arXiv:gr-qc/9310026.

\bibitem{Jacobson1995}
T.~Jacobson.
\newblock Thermodynamics of spacetime: The {E}instein equation of state.
\newblock {\em Phys. Rev. Lett.}, 75:1260, 1995.

\bibitem{Susskind1995}
L.~Susskind.
\newblock The world as a hologram.
\newblock {\em J. Math. Phys.}, 36:6377, 1995.

\bibitem{Bousso2002}
R.~Bousso.
\newblock The holographic principle.
\newblock {\em Rev. Mod. Phys.}, 74:825, 2002.

\bibitem{Verlinde2011}
E.~Verlinde.
\newblock On the origin of gravity and the laws of {N}ewton.
\newblock {\em JHEP}, 1104:029, 2011.

\bibitem{CohenKaplanNelson1999}
A.~G. Cohen, D.~B. Kaplan, and A.~E. Nelson.
\newblock Effective field theory, black holes, and the cosmological constant.
\newblock {\em Phys. Rev. Lett.}, 82:4971, 1999.

\bibitem{Hogan2014}
C.~J. Hogan.
\newblock Quantum entanglement of matter and geometry in large systems.
\newblock arXiv:1412.1807 [gr-qc], 2014.

\bibitem{Hogan2008a}
C.~J. Hogan.
\newblock Measurement of quantum fluctuations in geometry.
\newblock {\em Phys. Rev. D}, 77:104031, 2008.

\bibitem{Hogan2008}
C.~J. Hogan.
\newblock Indeterminacy of holographic quantum geometry.
\newblock {\em Phys. Rev. D}, 78:087501, 2008.

\bibitem{Hogan2012}
C.~J. Hogan.
\newblock Interferometers as probes of {P}lanckian quantum geometry.
\newblock {\em Phys. Rev. D}, 85:064007, 2012.

\bibitem{KwonHogan2014}
O.~Kwon and C.~J. Hogan.
\newblock Interferometric tests of {P}lanckian quantum geometry models.
\newblock {\em Class. Quant. Grav.}, 33:105004, 2016.

\bibitem{Adhikari2014}
R.~X. Adhikari.
\newblock Gravitational radiation detection with laser interferometry.
\newblock {\em Rev. Mod. Phys.}, 86:121, 2014.

\bibitem{Hogan2012b}
C.~J. Hogan.
\newblock A model of macroscopic geometrical uncertainty.
\newblock arXiv:1204.5948 [gr-qc], 2012.

\bibitem{AffeldtEtAl2014}
C.~Affeldt et~al.
\newblock Advanced techniques in {GEO} 600.
\newblock {\em Class. Quant. Grav.}, 31:224002, 2014.

\bibitem{DooleyLIGO2015}
{K. L. Dooley for the {LIGO} Scientific Collaboration}.
\newblock Status of {GEO} 600.
\newblock {\em J. Phys. Conf. Ser.}, 610:012015, 2015.

\bibitem{ChouEtal2015}
{A. Chou et al. (Fermilab Holometer Collaboration)}.
\newblock First measurements of high frequency cross-spectra from a pair of
  large {M}ichelson interferometers.
\newblock {\em Phys. Rev. Lett.}, 117:111102, 2016.

\bibitem{ChouEtal:2016}
{A. Chou et al. (Fermilab Holometer Collaboration)}.
\newblock The {H}olometer: An instrument to probe {P}lanckian quantum geometry.
\newblock {\em Class. Quant. Grav.}, 34:065006, 2017.

\bibitem{LIGO2015a}
{The {LIGO} Scientific Collaboration}.
\newblock {A}dvanced {LIGO}.
\newblock {\em Class. Quant. Grav.}, 32:074001, 2015.

\bibitem{LIGO2016}
{The {LIGO} Scientific Collaboration} and {The Virgo Collaboration}.
\newblock {GW150914}: The {A}dvanced {LIGO} detectors in the era of first
  discoveries.
\newblock {\em Phys. Rev. Lett.}, 116:131103, 2016.

\bibitem{Hogan2015}
C.~J. Hogan.
\newblock Exotic rotational correlations in quantum geometry.
\newblock arXiv:1509.07997 [gr-qc], 2015.

\bibitem{HoganKwonRichardson2016}
C.~J. Hogan, O.~Kwon, and J.~Richardson.
\newblock Statistical model of exotic rotational correlations in emergent
  space-time.
\newblock arXiv:1607.03048 [gr-qc], 2016.

\bibitem{SaleckerWigner1958}
H.~Salecker and E.~P. Wigner.
\newblock Quantum limitations of the measurement of space-time distances.
\newblock {\em Phys. Rev.}, 109:571, 1958.

\bibitem{Caves1980}
C.~M. Caves.
\newblock Quantum-mechanical radiation-pressure fluctuations in an
  interferometer.
\newblock {\em Phys. Rev. Lett.}, 45:75, 1980.

\bibitem{CavesThorneDreverSandbergZimmermann1980}
C.~M. {Caves}, K.~S. {Thorne}, R.~W.~P. {Drever}, V.~D. {Sandberg}, and
  M.~{Zimmermann}.
\newblock On the measurement of a weak classical force coupled to a
  quantum-mechanical oscillator. {I}. {I}ssues of principle.
\newblock {\em Rev. Mod. Phys.}, 52:341, 1980.

\bibitem{RuoBercheraDegiovanniOlivaresGenovese2013}
I.~{Ruo Berchera}, I.~P. {Degiovanni}, S.~{Olivares}, and M.~{Genovese}.
\newblock Quantum light in coupled interferometers for quantum gravity tests.
\newblock {\em Phys. Rev. Lett.}, 110:213601, 2013.

\bibitem{Siegman1986lasers}
A.~E. Siegman.
\newblock {\em Lasers}.
\newblock University Science Books, Sausalito, 1986.

\bibitem{Gardiner2004quantum}
C.~Gardiner and P.~Zoller.
\newblock {\em Quantum Noise: A Handbook of Markovian and Non-Markovian Quantum
  Stochastic Methods with Applications to Quantum Optics}.
\newblock Springer Series in Synergetics. 2004.

\bibitem{OppenheimVerghese2015}
A.~V. Oppenheim and G.~C. Verghese.
\newblock {\em Signals, Systems and Inference}.
\newblock Pearson Education, 2015.
\newblock Chapter 11.

\bibitem{NgDam2000}
Y.~J. Ng and H.~van Dam.
\newblock Measuring the foaminess of space-time with gravity-wave
  interferometers.
\newblock {\em Found. Phys.}, 30:795, 2000.

\bibitem{Ng2002}
Y.~J. Ng.
\newblock Spacetime foam.
\newblock {\em Int. J. Mod. Phys. D}, 11:1585, 2002.

\bibitem{FrolovGrote2014}
V.~Frolov and H.~Grote.
\newblock Gravitational wave detectors in {E}urope and the {U.S.}
\newblock Fermilab Seminar, {April 7,} 2014.
\newblock URL:
  \url{http://astro.fnal.gov/wp-content/uploads/2014/09/VFrolov_HGrote-040714.pdf}.

\bibitem{Hild2009}
{S. Hild for the GEO-600 Collaboration}.
\newblock Recent experiments in {GEO600} regarding the holographic noise
  hypothesis.
\newblock Holographic Noise Workshop, Albert Einstein Institute, May 2009.
\newblock URL:
  \url{http://www.physics.gla.ac.uk/~shild/presentations/holographic_noise_experiments_GEO600.pdf}.

\bibitem{ChouWeiss2009}
{A. Chou, R. Weiss et al. (Fermilab Holometer Collaboration)}.
\newblock The {F}ermilab {H}olometer: A program to measure {P}lanck scale
  indeterminacy.
\newblock 2009.
\newblock URL:
  \url{http://www.fnal.gov/directorate/program_planning/Nov2009PACPublic/holometer-proposal-2009.pdf}.

\bibitem{Kamai2013}
{B. Kamai et al. (Fermilab Holometer Collaboration)}.
\newblock The {F}ermilab {H}olometer: Probing the {P}lanck scale.
\newblock In {\em Am. Astron. Soc. Meeting \#221 \#431.06}, 2013.

\bibitem{LieuHillman2003}
R.~Lieu and L.~W. Hillman.
\newblock The phase coherence of light from extragalactic sources: Direct
  evidence against first-order {P}lanck-scale fluctuations in time and space.
\newblock {\em Astrophys. J.}, 585:L77, 2003.

\bibitem{RagazzoniTurattoGaessler2003}
R.~Ragazzoni, M.~Turatto, and W.~Gaessler.
\newblock The lack of observational evidence for the quantum structure of
  spacetime at {P}lanck scales.
\newblock {\em Astrophys. J.}, 587:L1, 2003.

\bibitem{NgChristiansenDam2003}
Y.~J. Ng, W.~A. Christiansen, and H.~van Dam.
\newblock Probing {P}lanck-scale physics with extragalactic sources?
\newblock {\em Astrophys. J.}, 591:L87, 2003.

\bibitem{ChristiansenNgDam2006}
W.~A. Christiansen, Y.~J. Ng, and H.~van Dam.
\newblock Probing spacetime foam with extragalactic sources.
\newblock {\em Phys. Rev. Lett.}, 96:051301, 2006.

\bibitem{ChristiansenNgFloydPerlman2011}
W.~A. Christiansen, Y.~J. Ng, D.~J.~E. Floyd, and E.~S. Perlman.
\newblock Limits on spacetime foam.
\newblock {\em Phys. Rev. D}, 83:084003, 2011.

\bibitem{PerlmanNgFloydChristiansen2011}
E.~S. Perlman, Y.~J. Ng, D.~J.~E. Floyd, and W.~A. Christiansen.
\newblock Using observations of distant quasars to constrain quantum gravity.
\newblock {\em Astronom. and Astrophys.}, 535:L9, 2011.

\bibitem{PerlmanRappaportChristiansenNg2015}
E.~S. Perlman, S.~A. Rappaport, W.~A. Christiansen, Y.~J. Ng, J.~DeVore, and
  D.~Pooley.
\newblock New constraints on quantum gravity from {X}-ray and {G}amma-ray
  observations.
\newblock {\em Astrophys. J.}, 805:10, 2015.

\end{thebibliography}

\newpage

\appendix
%dummy comment inserted by tex2lyx to ensure that this paragraph is not empty%dummy comment to ensure that this paragraph is not empty

\section*{Appendix:  Normalization from Spin Algebra and Holographic Gravity}

It is useful
to adopt an exact benchmark value for a Planck scale in absolute physical units,  based
on the states of a particular quantum system that has both a precisely calculable
information content and a precisely calculable position variance.
We adopt here the following simple model of a quantum system based
on the familiar spin algebra that has these properties, and does not
depend on the system dynamics. Of course, just because this model
is precise does not mean it is correct. This particular way of counting
information\cite{Hogan2012b} is not a necessary assumption for the
validity of the generalized framework proposed above, which relies
only on first principles such as causal structure and symmetries of
spacetime.

Consider a noncommutative geometry described by a standard spin algebra in three dimensions $i,j,k=1,2,3$: 
\begin{equation}
[{\hat{x}}_{i},{\hat{x}}_{j}]={\hat{x}}_{k}\epsilon_{ijk}i\ell_{P}.\label{3Dcommute}
\end{equation}
where $\epsilon_{ijk}$ denotes the antisymmetric tensor.
Consider an operator 
\begin{equation}
|{\hat{x}}|^{2}\equiv{\hat{x}}_{i}{\hat{x}}_{i},
\end{equation}
corresponding to the squared modulus of separation, analogous to the
square of total angular momentum. Using conventional notation used for angular momentum, let $l$
denote positive integers corresponding to the quantum numbers of radial
separation, analogous to total angular momentum. The separation operator
takes discrete eigenvalues: 
\begin{equation}
|{\hat{x}}|^{2}|l\rangle=l(l+1)\ell_{P}^{2}|l\rangle.\label{radialeigenvalues}
\end{equation}
We denote the discrete eigenvalues corresponding to classical separation
by 
\begin{equation}
L\equiv\sqrt{l(l+1)}\ell_{P}.\label{separation}
\end{equation}
It can be shown that position ${\hat{x}}_{\perp}$ in any direction transverse
to separation is indeterminate, with a variance given by 
\begin{equation}
\langle{\hat{x}}_{\perp}^{2}\rangle=L\ell_{P}.\label{perpvariance}
\end{equation}
Like spin, the states have a discrete spectrum. The number
of position eigenstates within a 3-sphere of radius $R$ can be counted
in the same way as discrete angular momentum eigenstates. Using
Eq. (\ref{separation}) with $R=L$, the number $l_{R}$ of radial
position eigenstates for a radius $R$ is given by setting $l_{R}(l_{R}+1)=(R/\ell_{P})^{2}$.
For each of these, there are
$2l+1$ eigenstates of direction. The total number of quantum position
eigenstates in a 3-sphere is then 
\begin{equation}
{\cal N}_{Q3S}(R)=\sum_{l=1}^{l_{R}}(2l+1)=l_{R}(l_{R}+2)=(R/\ell_{P})^{2},\label{3Dsphere}
\end{equation}
where the last equality applies in the large $l$ limit. Thus, the
number of quantum-geometrical position eigenstates in a volume scales
holographically, as the surface area in Planck units.

We can set an absolute physical scale based on equating  the number of states in a 3-sphere from 
spin geometry and from 
thermodynamics of gravitational systems\cite{Jacobson1995,Verlinde2011}.
The number of position states for a massive body enclosed in a 3-sphere
of radius $R$ that statistically reproduces nonrelativistic Newtonian
gravity (cf. \cite{Verlinde2011}, Eq. 3.10) is given by: 
\begin{equation}
{\cal N}_{G3S}(R)=4\pi(R/ct_{P})^{2},\label{verlindegravity}
\end{equation}
with the usual definition of Planck length, $ct_{P}\equiv\sqrt{\hbar G/c^{3}}=1.616\times10^{-35}$m.
Note that this is four times larger than the entropy of a black hole
event horizon of the same radius. The quantum-geometrical and gravitational
estimates agree, ${\cal N}_{Q3S}={\cal N}_{G3S}$, if the numerical
value of the effective commutator coefficient $\ell_{P}$ is 
\begin{equation}
\ell_{P}=ct_{P}/\sqrt{4\pi}.\label{value}
\end{equation}
If Eq. (\ref{value}) holds, bodies in the emergent space-time move
as described by Verlinde's entropic Newtonian gravity.
From Eq. (\ref{perpvariance}), the theory predicts position variance
in physical units, 
\begin{equation}
\langle{\hat{x}}_{\perp}^{2}\rangle=Lct_{P}/\sqrt{4\pi}=(2.135\times10^{-18}{\rm m})^{2}(L/{\rm 1m}),\label{exact}
\end{equation}
with no free parameters. 
A less conservative normalization is to use the holographic principle directly, and equate the number of states in the sphere with those in an event horizon of the same Schwarzschild radius. That  gives a larger value, $\ell_{P}'=ct_{P}/\sqrt{\pi}$, so it is easier to rule out experimentally.
\end{document}